\documentclass[twocolumn,english]{article}
\usepackage[T1]{fontenc}
\usepackage[latin9]{inputenc}
\usepackage{geometry}
\usepackage{float}
\usepackage{rotfloat}
\usepackage{amsmath}
\usepackage{amssymb}
\usepackage{graphicx}
\usepackage{setspace}
\usepackage[authoryear]{natbib}

\makeatletter

\providecommand{\tabularnewline}{\\}
\floatstyle{ruled}
\newfloat{algorithm}{tbp}{loa}
\providecommand{\algorithmname}{Algorithm}
\floatname{algorithm}{\protect\algorithmname}

\makeatother

\usepackage{babel}
\begin{document}


\title{Modified Cholesky Riemann Manifold Hamiltonian Monte Carlo: Exploiting
Sparsity for Fast Sampling of High-dimensional Targets\thanks{The author would like to thank the Editor, the Associate Editor, two referees, 
Michael Betancourt, Hans J. Skaug and Anders Tranberg for comments that have sparked many 
improvements.}}


\author{Tore Selland Kleppe 
\footnote{Department of Mathematics and Natural Sciences, University of Stavanger,
4036 Stavanger, Norway.  Telephone: +47
95922364. tore.kleppe@uis.no}}

\maketitle 
\begin{abstract}
Riemann manifold Hamiltonian Monte Carlo (RMHMC) has the potential
to produce high-quality Markov chain Monte Carlo-output even for
very challenging target distributions. To this end, a symmetric positive
definite scaling matrix for RMHMC, which derives, via a modified Cholesky
factorization, from the potentially indefinite negative Hessian of
the target log-density is proposed. The methodology is able to exploit
the sparsity of the Hessian, stemming from conditional independence modelling assumptions,
and thus admit fast implementation
of RMHMC even for high-dimensional target distributions. Moreover,
the methodology can exploit log-concave conditional target densities,
often encountered in Bayesian hierarchical models, for faster sampling
and more straight forward tuning. The proposed methodology is compared
to alternatives for
some challenging targets, and is illustrated by applying a state space
model to real data.
\end{abstract}

\section{Introduction}

Markov chain Monte Carlo (MCMC) methods have by now seen widespread
use for sampling from otherwise intractable distributions in statistics
for close to three decades \citep{GelmanBDA3}. Still the development
of new and improved MCMC methods for tackling ever more challenging
sampling problems is a highly active field \citep[see e.g.][]{RSSB:RSSB736,girolami_calderhead_11,Calderhead09122014,JMLR:v15:hoffman14a}.
The contribution of the present work is a new metric tensor, deriving
directly from the Hessian of the log-target density that, together
with Riemann manifold Hamiltonian Monte Carlo (RMHMC) \citep{girolami_calderhead_11}
enables fast and robust sampling from target distributions with strong
non-linear dependencies. Such target distributions arise, for instance,
as the joint posterior distribution of latent variables and parameters
in non-linear and/or non-Gaussian Bayesian hierarchical models.

Current MCMC strategies for Bayesian inference in such hierarchical
models can informally be split into three categories \citep[see also][for a similar discussion]{1312.0906}:
1) Variants of Gibbs sampling, 2) Pseudo-marginal methods and 3) methods
that update latent variables and parameters jointly. Gibbs sampling
\citep[see e.g.][]{liu_mc_2001,robert_casella} is widely used as
it is, in many cases, relatively easy to implement. However, it is well
known that naive Gibbs sampling for hierarchical models, where e.g. the latent variables are in
one block and the variance parameter of the latent variables is in
another block, can lead to poor mixing
due to strong non-linear dependencies across the blocks. 

Pseudo-marginal methods \citep[see e.g.][]{RSSB:RSSB736,Pitt2012134}
on the other hand, seek to avoid such poor mixing by targeting directly the
marginal posterior of the parameters (i.e. with the latent variables
integrated out). However, such methods
hinge on the ability to Monte Carlo simulate an unbiased, low variance
estimate of marginal posterior density of the parameters, which can often
be extremely computationally demanding \citep{flury_shephart_2011},
or even infeasible for larger models. 

Finally, methods that update the latent variables and parameters jointly
are attractive in theory as they also avoid the non-linear dependency
problems of Gibbs sampling, and they do not need computationally demanding
marginal density estimates. However, such methods need mechanisms
for aligning the proposals with the local geometry of the target,
in particular for high-dimensional problems. Currently, popular methods
within this category are Metropolis-adjusted Langevin methods \citep[see e.g.][]{roberts_stramer_02,girolami_calderhead_11,Xifara201414,SJOS:SJOS12204}
and various variants of Hamiltonian Monte Carlo 
\citep[see e.g.][]{Duane1987216,neal_thesis,1206.1901,girolami_calderhead_11,1212.4693,doi:10.1080/10618600.2014.902764}.
Both use derivative information from the target log-density
to guide the MCMC proposals. However, it has become clear \citep[see e.g.][]{1212.4693}
that methods based on first order derivatives only (as is done in the default
MCMC method in the popular Bayesian computation software Stan \citep{JSSv076i01}), 
can be inefficient
if one fails to take the local
scaling properties of the target into account. This mirrors the relation between the 
method of steepest descent and Newton's method in numerical
optimisation \citep{noce:wrig:1999}. 

Joint updating of latent variables and parameters in Bayesian hierarchical
models is also the main motivation of this paper, even though the
methodology is applicable for any continuous target distribution
under some regularity conditions. Here a particular modified
Cholesky factorization is proposed, that when applied to a potentially indefinite
negative Hessian of the log-target, produces useful scaling information
for RMHMC. In conjunction, RMHMC and the proposed methodology enables
MCMC sampling where the proposals are far from the current configuration,
and that is robust to significantly different scaling properties across
the support of the target, a property often seen in Bayesian hierarchical
models. It is worth noticing that applying modified negative Hessians
in RMHMC \citep{1212.4693} or when scaling MCMC proposals in general
\citep{doi:10.1080/03610919908813583,doi:10.1081/STA-120018828,qi_minka,martin_etal2012,SJOS:SJOS12204}
is not new per se. However, the proposed modified Cholesky approach
is, by exploiting sparsity of the negative Hessian, computationally
fast and scalable in the dimension of the target. 

The exploitation of sparsity is increasingly important in the numerical linear 
algebra involved in modern statistical computing associated with hierarchical models, 
as the dimension of matrices to be factorized can easily reach $\sim10^{5}$
\citep[see e.g.][]{RSSB:RSSB288,rue_held_05,RSSB:RSSB700}. The sparsity of involved Hessian
matrices arises due to conditional 
independence assumptions used in the modelling. Examples include block-diagonal structures 
associated with non-linear mixed effect regressions, banded structures for Markovian dynamic 
models with unobserved factors such as state space models, and less structured sparse matrices for 
spatial/spatial-temporal models that involve Gaussian Markov random fields \citep[see e.g.][]{RePEc:bla:jorssb:v:73:y:2011:i:4:p:423-498}. 
Currently, the Integrated Nested Laplace Approximation (INLA) \citep{RSSB:RSSB700} 
is a widely used methodology for fast approximate Bayesian inference in 
the Latent Gaussian Model (LGM) sub-class of
Bayesian hierarchical models. Like the proposed methodology, INLA relies heavily on
exploiting sparsity in order to speed up computations, and in the context of
LGMs, the proposed methodology can also benefit from the fact that conditional posterior log-densities
of the latents are concave. However, the proposed methodology is more general with respect
to models that can be handled, and in particular does not require the LGM-assumption that
the latent variables have a joint Gaussian prior (see Section 5), or the INLA-assumption 
that the number of parameters is small.

The remainder of the paper is laid out as follows: Section 2 fixes
notation and reviews RMHMC. Section 3 describes and discusses the
proposed methodology. In Section 4, the proposed methodology is compared
to Gibbs sampling, Euclidian metric Hamiltonian Monte Carlo (EHMC), The no-u-turn sampler (NUTS)
of Stan and RMHMC based on spectral decompositions 
for two challenging target distributions. Section 5 describes an application
to a non-linear, non-Gaussian state space model, and finally Section
6 provides some discussion.

\section{Riemann manifold HMC}

This section fixes notation and reviews RMHMC \citep{girolami_calderhead_11} in order to set
the stage. Denote by $\nabla_{\mathbf{y}}$ the gradient/Jacobian
operator with respect to vector $\mathbf{y}$, $|A|$ the determinant
of square matrix $A$, and $I_{d}$ denotes the $d$-dimensional identity
matrix. A natural matrix norm is denoted by $\Vert\cdot\Vert$ and
$\Phi$ is the standard normal cumulative distribution function.

Let $\tilde{\pi}(\mathbf{x})$ denote a density kernel associated
with the target density $\pi(\mathbf{x}):\Omega\rightarrow\mathbb{R}^{+}$
where $\Omega\subseteq\mathbb{R}^{d}$. It is assumed that $\tilde{\pi}(\mathbf{x})$
is continuous and has continuous derivatives up to order 3. Moreover,
let the metric tensor $G(\mathbf{x})$ be a symmetric positive definite
$d\times d$ matrix for all $\mathbf x \in \Omega$, where $G_{i,j}(\mathbf{x}):\Omega\rightarrow\mathbb{R}$
are smooth functions for all $i,j=1,\dots,d$. Particular choices
of $G(\mathbf{x})$ will be discussed in detail in Section \ref{sec:The-regularized-Newton}. 

Like for other Hamiltonian Monte Carlo methods, RMHMC relies on defining
a synthetic Hamiltonian dynamical system that evolves over fictitious
time $\tau$, where $\mathbf{x}$ plays the role of position variable
and $\mathbf{p}\in\mathbb{R}^{d}$ is the (auxiliary) momentum variable.
The total energy in the system is given by the Hamiltonian $H(\mathbf{x},\mathbf{p})$,
which for RMHMC is taken to be 
\begin{equation}
H(\mathbf{x},\mathbf{p})=-\log\tilde{\pi}(\mathbf{x})+\frac{1}{2}\log(|G(\mathbf{x})|)+\frac{1}{2}\mathbf{p}^{T}G(\mathbf{x})^{-1}\mathbf{p}.\label{eq:hamiltonian}
\end{equation}
The time-evolution of $(\mathbf{x}(\tau),\mathbf{p}(\tau))$ is described
by Hamilton's equations
\begin{eqnarray}
\frac{\partial}{\partial\tau}\mathbf{x}(\tau) & = & \nabla_{\mathbf{p}}H(\mathbf{x}(\tau),\mathbf{p}(\tau))=G(\mathbf{x}(\tau))^{-1}\mathbf{p(\tau)},\label{eq:x_evol}\\
\frac{\partial}{\partial\tau}\mathbf{p}(\tau) & = & -\nabla_{\mathbf{x}}H(\mathbf{x}(\tau),\mathbf{p}(\tau)).\label{eq:p_evol}
\end{eqnarray}
Let $\mathbf{z}(\tau)=(\mathbf{x}(\tau)^T,\mathbf{p}(\tau)^T)^T \in\mathbb{R}^{2d}$
be the state of the system at time $\tau$ and likewise define $H(\mathbf{z}(\tau))=H(\mathbf{x}(\tau),\mathbf{p}(\tau))$.
Moreover, let $\varphi_{\tau}(\cdot):\mathbb{R}\times\Omega\rightarrow\Omega$
denote the flow of $\mathbf{z}$ associated with (\ref{eq:x_evol},\ref{eq:p_evol})
so that $\mathbf{z}(r+s)=\varphi_{s}(\mathbf{z}(r))\;\forall r,s\in\mathbb{R}$
whenever $\mathbf{z}(\tau)$ solves (\ref{eq:x_evol},\ref{eq:p_evol}).
The following properties of the Hamiltonian flow $\varphi_{\tau}$
can be established \citep[see e.g.][]{Leimkuhler:2004} 
\begin{itemize}
\item Energy conservation, i.e. $H(\varphi_{\tau}(\mathbf{z}))$ is constant
as a function of $\tau$ for any $\mathbf{z}\in\Omega$.
\item Time-reversibility, i.e. the inverse of $\varphi_{\tau}$ is $\varphi_{-\tau}$
so that $\varphi_{-\tau}(\varphi_{\tau}(\mathbf{z}))=\mathbf{z}$
for any $\mathbf{z}\in\Omega$.
\item $\varphi_{\tau}$ is said to be symplectic, namely for each $\tau$,
$(\nabla_{\mathbf{z}}\varphi_{\tau}(\mathbf{z}))^{T}J(\nabla_{\mathbf{z}}\varphi_{\tau}(\mathbf{z}))=J$
where 
\[
J=\left[\begin{array}{cc}
0 & I_{d}\\
-I_{d} & 0
\end{array}\right].
\]
In particular, the symplecticity implies that $\varphi_{\tau}$ is
a volume-preserving map so that the Jacobian $\nabla_{\mathbf{z}}\varphi_{\tau}(\mathbf{z})$
has unit determinant. 
\end{itemize}
Based on these properties, it is relatively straight forward to verify
that $\varphi_{\tau}$ preserves the Bolzmann distribution
\[
\pi(\mathbf{z})=\pi(\mathbf{x},\mathbf{p})\propto\exp(-H(\mathbf{x},\mathbf{p})).
\]
Namely, provided that $\mathbf{z}\sim\pi(\mathbf{z})$, then also
$\varphi_{\tau}(\mathbf{z})\sim\pi(\mathbf{z})$ for any $\tau\in\mathbb{R}$.
Given the particular specification of the Hamiltonian (\ref{eq:hamiltonian}),
the Bolzmann distribution admits the target distribution $\pi(\mathbf{x})$
as the $\mathbf{x}$-marginal. To see this, observe that
\begin{multline}
\int\pi(\mathbf{x},\mathbf{p})d\mathbf{p} \\
\propto\hat{\pi}(\mathbf{x})|G(\mathbf{x})|^{-\frac{1}{2}}
\int\exp\left(-\frac{1}{2}\mathbf{p}^{T}G(\mathbf{x})^{-1}\mathbf{p}\right)d\mathbf{p}\\
\propto\hat{\pi}(\mathbf{x}).
\end{multline}

Granted the above constructions, an ideal MCMC algorithm for obtaining
(dependent) samples $\{(\mathbf{x}_{t},\mathbf{p}_{t})\}_{t}\sim\pi(\mathbf{x},\mathbf{p})$
would be to alternate between 1) sample $\mathbf{p}_{t-1}\sim\pi(\mathbf{p}|\mathbf{x}_{t-1})=N(0,G(\mathbf{x}_{t-1}))$,
and 2) compute $(\mathbf{x}_{t},\mathbf{p}^{*})=\varphi_{\tau}((\mathbf{x}_{t-1},\mathbf{p}_{t-1}))$
for some $\tau$. However, for most non-trivial target distributions
and metric tensors, such an algorithm is infeasible as the corresponding $\varphi_{\tau}$s
do not admit closed form expressions. Instead, RMHMC relies on approximate
numerical simulation of the flow and correcting for the numerical
error using an accept-reject step.

\subsection{Numerical simulation of the flow and RMHMC}

In order to simulate the flow numerically, the splitting method
integrator of \citet{1212.4693} was used. This integrator preserves the time-reversibility
and symplectic nature of the flow, while only approximately preserves
the total energy. Though alternative, potentially less computationally
intensive, non-symplectic implementations exist \citep[see e.g.][]{doi:10.1080/10618600.2014.902764},
a symplectic integrator was chosen, as it admits stable and accurate numerical
simulation of the flow for large $d$ and potentially long time spans.
Moreover, using a symplectic and time-reversible integrator leads
to simple expressions for the accept probability used in the accept/reject
step. 

The integrator of \citet{1212.4693} for approximating $\varphi_{\varepsilon}(\mathbf{z}(\tau))$,
for some small time step $\varepsilon$, is characterized by 
\begin{multline}
\hat{\mathbf{p}}_{*}  =  \mathbf{p}(\tau)\\
-\frac{\varepsilon}{2}\nabla_{\mathbf{x}}\left[-\log\tilde{\pi}(\mathbf{x}(\tau))+\frac{1}{2}\log(|G(\mathbf{x}(\tau))|)\right],\label{eq:p_update1}
\end{multline}
\begin{equation}
\hat{\mathbf{p}}_{**}  =  \hat{\mathbf{p}}_{*}-\frac{\varepsilon}{2}\nabla_{\mathbf{x}}\left[\frac{1}{2}\mathbf{\hat{p}_{**}}^{T}G(\mathbf{x}(\tau))^{-1}\hat{\mathbf{p}}_{**}\right],\label{eq:implicit_1}
\end{equation}
\begin{multline}
\hat{\mathbf{x}}(\tau+\varepsilon)  =  \mathbf{x}(\tau)+\frac{\varepsilon}{2}G^{-1}(\mathbf{x}(\tau))\hat{\mathbf{p}}_{**}\\
+\frac{\varepsilon}{2}G^{-1}(\hat{\mathbf{x}}(\tau+\varepsilon))\hat{\mathbf{p}}_{**},\label{eq:implicit_2}
\end{multline}
\begin{equation}
\hat{\mathbf{p}}(\tau+\varepsilon)  =  \hat{\mathbf{p}}_{**}-\frac{\varepsilon}{2}\nabla_{\mathbf{x}}H(\mathbf{x}(\tau+\varepsilon),\hat{\mathbf{p}}_{**}),\label{eq:GL_explicit}
\end{equation}
where $\hat{\mathbf{x}}(\tau)$ and $\hat{\mathbf{p}}(\tau)$ denote
the approximations to $\mathbf{x}(\tau)$ and $\mathbf{p}(\tau)$
respectively. Applying the integrator (\ref{eq:p_update1}-\ref{eq:GL_explicit})
sequentially $l=1,2,\dots$ times produces approximations to the flow
after $\varepsilon l$ time has passed. A single transition $\mathbf{x}_{t-1}\rightarrow\mathbf{x}_{t}$
of the basic RMHMC algorithm used throughout this paper can be
summarized by the steps:
\begin{enumerate}
\item Resample momentums $\mathbf{p}_{t-1}\sim\pi(\mathbf{p}|\mathbf{x}_{t-1})=N(0,G(\mathbf{x}_{t-1}))$.
\item Sample $l$ and $\varepsilon$, and perform $l$ integrator steps with step size $\varepsilon$
starting at $(\text{\ensuremath{\mathbf{x}}}_{t-1},\mathbf{p}_{t-1})$.
This process results in the proposal $(\mathbf{x}_{t}^{*},\mathbf{p}_{t}^{*})$.
\item With probability $\min(1,\exp(-H(\mathbf{x}_{t}^{*},\mathbf{p}_{t}^{*})+H(\text{\ensuremath{\mathbf{x}}}_{t-1},\mathbf{p}_{t-1})))$
set $\mathbf{x}_{t}=\mathbf{x}_{t}^{*}$, and with remaining probability
set $\mathbf{x}_{t}=\mathbf{x}_{t-1}$.
\end{enumerate}
Other, more complicated and potentially more efficient variants of
the overarching RMHMC algorithm, such as algorithms choosing the number of integration
steps dynamically \citep{1304.1920,1601.00225}
are also conceivable in this framework. Such dynamic selection is likely to be required
in more automated implementations of the proposed methodology. However,
the basic RMHMC method outlined above was used, as the
focus of the present paper is on the particular metric tensor advocated. 

It is worth noticing that (\ref{eq:implicit_1},\ref{eq:implicit_2})
are implicit and therefore necessitate computationally costly fixed
point iterations \citep{Leimkuhler:2004}. In fact, the bulk part of the computation is spent
computing the derivatives of $H$ needed for solving (\ref{eq:implicit_1},\ref{eq:implicit_2}),
and therefore implementing the solution process in an efficient manner
is of high importance. In the present implementation, the fixed point iterations
are continued until the infinity-norm of the difference between successive
iterates is $<1.0e-6$. To meet this tolerance in the real application with $d=3087$ considered 
in Section 5, 5-6 iterations are required in (\ref{eq:implicit_1}) and 3-4 iterations are required in
(\ref{eq:implicit_2}).

\section{Metric tensor based on a modified Cholesky factorization\label{sec:The-regularized-Newton}}

This section describes the proposed metric tensor, and thereby the implied
Riemann manifold intended for RMHMC and related methods.
By now, a rich literature considering the differential geometric properties
of RMHMC has appeared \citep[see e.g.][]{1410.5110}. In this paper,
a slightly less mathematically inclined approach is taken, and rather
focus lies on some intuition and how to implement RMHMC for a general target
distribution. 
\subsection{Metric tensor from the negative Hessian}
For EHMC, a rule of thumb is that a constant $G$ should
be taken as the precision matrix of the target for near-Gaussian targets
\citep{1206.1901}. For RMHMC, several papers have argued for a metric
tensor deriving from (negative) second derivative information, specifically
the Fisher information matrix \citep{girolami_calderhead_11,doi:10.1080/10618600.2014.902764}
or a regularized version of the negative log-target density Hessian
\citep[see e.g.][]{disc_sanzserna,disc_NTNU,disc_Jasra}.
Modulus regularisation, the latter
fulfils the rule of thumb for EHMC and Gaussian targets. For a model
where the Fisher information matrix is available, it is likely that
the information matrix approach is preferable, as the information matrix is by construction
positive definite. However, for a general model, the information matrix
is either not available in closed form or requires substantial analytic calculations
for each model instance. In the reminder of the paper, unavailable Fisher information
matrix is taken as a premise for the discussion.

In deriving the metric tensor directly from
the negative Hessian, computing expectations is no longer needed, but on the
other hand, accounting for the fact that the negative Hessian often is indefinite
in non-negligible subsets of $\Omega$ is required. \citet{1212.4693} introduced
the softmax metric, which is based on a full spectral decomposition
of the negative Hessian and a subsequent regularisation of the eigenvalues.
This method is attractive as it retains the eigenvectors of the negative
Hessian, but is very computationally demanding for large models.
The present work is similar to \citet{1212.4693} in deriving $G$
directly from the Hessian, but the computational details are different. 

An additional rationale for choosing $G$ to be a regularized approximation
to the negative Hessian is as follows. It is relatively straight forward
to verify that a RMHMC proposal for $\mathbf{x}$ using a single time
integration step (using the generalized leapfrog integrator \citep[page 156]{girolami_calderhead_11,Leimkuhler:2004}),
starting at $\mathbf{x}(0)$ will have the mean 
\begin{multline}
E(\hat{\mathbf{x}}(\varepsilon)|\mathbf{x}(0))  =  \mathbf{x}(0)\\
+\frac{\varepsilon^{2}}{2}\left[G^{-1}(\mathbf{x}(0))\left[\nabla_{\mathbf{x}}\log\tilde{\pi}(\mathbf{x}(0))\right]+\Lambda(\mathbf{x}(0))\right]\\
+O(\varepsilon^{4}),\label{eq:RMHMC_one_step_mean}
\end{multline}
where
\[
\Lambda_{i}(\mathbf{x})  = \sum_{j=1}^{d}\frac{\partial}{\partial x_{j}}G_{i,j}^{-1}(\mathbf{x}),\;i=1,\dots,d.
\]
It is seen that the $O(\varepsilon^{2})$ term scales the
gradient of the log-target with the inverse of $G$. Therefore choosing
$G$ as a regularized version of the negative Hessian will turn the
former part of the $O(\varepsilon^{2})$ term into a modified Newton
(numerical optimization) direction, which is known to be a well-scaled
search direction in the numerical optimization literature \citep[Chapter 6]{noce:wrig:1999}.
In fact, modulus the effect of the additional term $\Lambda$, choosing
$G$ in this manner makes time $\tau$ and the time step $\varepsilon$
effectively dimensionless \citep[Page 132]{girolami_calderhead_11}
as there is a unit ``natural'' step length \citep[Page 23]{noce:wrig:1999}
associated with a Newton step. In the present setup, this corresponds
to $\tau=O(\sqrt{2})$ is the time needed to traverse to the mode
from any region close to a mode. 

It is also worth noticing that the additional term $\Lambda$ in (\ref{eq:RMHMC_one_step_mean})
is a correction term that accounts for the non-constant curvature of
the implied manifold, while retaining the correct Boltzmann distribution
associated with (\ref{eq:hamiltonian}). In particular, the explicit
terms of (\ref{eq:RMHMC_one_step_mean}), along with the leading term
of 
\[
Var(\hat{\mathbf{x}}(\varepsilon)|\mathbf{x}(0))=\varepsilon^{2}G^{-1}(\mathbf{x}(0))+O(\varepsilon^{4}),
\]
are identical to the first two moments of the proposal
associated with the time discretised position dependent metric Langevin diffusion (with
$\pi(\mathbf{x})$ as stationary distribution due to the $\Lambda$
term in the drift) of \citet[Equation 9]{Xifara201414}. This observation
mirrors the well-known fact that one (Euclidian metric) Metropolis-adjusted
Langevin algorithm proposal is identical to the proposal of one EHMC
time integration step, but in the general Riemann manifold case, this
correspondence is only asymptotical as $\varepsilon\rightarrow0$.

At this point, it is also worth contrasting a variable $G(\mathbf x)$  to 
a constant $G$, resulting in EHMC. The latter is currently used (either identity, 
diagonal, or dense matrix) 
in tandem with the NUTS \citep{JMLR:v15:hoffman14a} as the 
default sampling algorithm in 
the widely applied Bayesian computation software Stan \citep{JSSv076i01}.
In the constant $G$-case, $\Lambda$ and the higher order terms of 
(\ref{eq:RMHMC_one_step_mean}) vanish, and thus the proposal means are
$E(\hat{\mathbf{x}}(\varepsilon)|\mathbf{x}(0))  =  \mathbf{x}(0)
+\frac{\varepsilon^{2}}{2}G^{-1}\left[\nabla_{\mathbf{x}}\log\tilde{\pi}(\mathbf{x}(0))\right]$.
For targets with close to constant curvature (i.e. in practice close to Gaussian, often seen
for posteriors in non-hierarchical models),
this works well, and is substantially faster than the proposed methodology as explicit
integrators may be employed. However, for targets exhibiting substantial variation
in the curvature, such fixed scaling of the gradient may, for non-negligible subsets of $\Omega$, 
produce either:
1) too aggressive proposals, leading to inaccurate simulation of the flow and subsequent rejections
of the proposals. In turn, this leads in practice to that subsets of $\Omega$ are left unexplored \citep{1212.4693}.
2) too defensive proposals, leading to slow exploration of subsets of $\Omega$, and in 
particular may often lead to too small $l$s being selected by NUTS-like algorithms \citep{1304.1920}.
In section 4, it is shown that Stan and EHMC may exhibit such pathologies when MCMC samples are compared to known
marginals of the target. However, in practice it is difficult to determine 
whether such pathologies are active in a given MCMC simulation.

The remainder of this section is devoted to the particular form of
metric tensor advocated here.

\subsection{A smooth modified Cholesky factorization}

This section develops a metric tensor in the form of a regularized
approximation of the negative Hessian matrix, which is guaranteed to be positive definite.
The approach is similar to the modified Cholesky factorization approach
for modified Newton optimization algorithms of \citet{gill_murray_1974,gilletal}.
The differences amount primarily to making sure that each element
of $G$ is a smooth function of $\mathbf{x}$ so that the implied
manifold is also smooth. Through a large number of numerical optimization
applications, the modified Cholesky approach has proven to be a trusted
and frequently used technology \citep{noce:wrig:1999}, and the approximation has  also
served as the basis for further refinements \citep{doi:10.1137/0911064,doi:10.1137/S105262349833266X}.
A further motivation for working with modified Cholesky factorizations,
as opposed to say methods based on full spectral decompositions, is
that this method can be implemented while exploiting sparsity patterns
\citep{davis_sparse} of the negative Hessian often found in statistical
models \citep[see e.g.][]{RSSB:RSSB288,RSSB:RSSB700}. 

Denote by $A$ a symmetric matrix (e.g. the negative Hessian) 
for which a positive definite approximation
is sought. The approach for finding $G$
takes as vantage point a square root free Cholesky factorization which
produces the (LDL-)decomposition 
\[
\bar{L}\bar{D}\bar{L}^{T}=A.
\]
Here, $\bar{L}$ is unit lower triangular (i.e. 1s on the diagonal)
and $\bar{D}$ is a diagonal matrix \citep[section 4]{golub_vanloan_3rd_ed}.
When $A$ is positive definite, the diagonal elements
of $\bar{D}$ are positive, whereas in the indefinite case, the factorization
may not exist or it may be numerically unstable and produce arbitrary large
elements in $\bar{L}$ and $\bar{D}$ \citep[][section 6.3]{noce:wrig:1999}.
The insight of \citet{gill_murray_1974,gilletal} was that $\bar{D}_{j,j}$
(and consequently the resulting $\bar{L}$) can be modified online
to produce a square root free Cholesky factorization $\tilde{L}D\tilde{L}=A+J$
of the \textit{symmetric positive definite} matrix $A+J$, where $J$ is a diagonal matrix
with non-negative diagonal elements. The diagonal elements of $J$ are chosen
by the modified Cholesky algorithm to be
large enough to ensure that $A+J$ is positive definite when $A$ is indefinite. On the other hand, 
when $A$ is sufficiently positive definite, $J$ is taken to be the zero-matrix
by the modified Cholesky algorithm. 
\begin{algorithm*}
\begin{tabbing}
\hspace{2em} \= \hspace{2em} \= \hspace{2em} \= \\
{\bfseries Input}: \\
\> -A $d\times d$ symmetric matrix $A$,\\
\> -A $d$-vector of regularisation parameters $\mathbf u$,\\
\> -{\bfseries Optionally} a positive integer $K\leq d$ indicating that $A_{1:K,1:K}$ is positive definite.\\
\> If no such information is present, then $K=0$.\\
step 0,1: $\tilde L \leftarrow I_d$.\\ 
step 0,2: $ D_{j,j} \leftarrow A_{j,j},\;j=1,\dots d$.\\
{\bfseries for} $j=1$ to $d$\\
\> step 1: {\bfseries if}$(j>1)$ $\tilde L_{j,k}\leftarrow \tilde L_{j,k}/ D_{k,k},\;k=1,\dots,j-1$. \\
\> step 2: {\bfseries if}$(j<d)$ $\tilde L_{j+1:d,j} \leftarrow  A_{j+1:d,j}$. \\
\> step 3: {\bfseries if}$(1<j<d)$ $\tilde L_{j+1:d,j} \leftarrow \tilde L_{j+1:d,j} -(\tilde L_{j+1:d,1:j-1})(\tilde L_{j,1:j-1})^T$.\\
\> step 4: {\bfseries if}$(j>K)$ $D_{j,j} \leftarrow \text{sabs}(D_{j,j};u_j)$. (See below for explanation of the smax function.)\\
\> step 5: {\bfseries if}$(j<d)$ $ D_{k,k} \leftarrow D_{k,k} - (\tilde L_{k,j})^2/ D_{j,j}, k=j+1,\dots,d$.\\
{\bfseries end for}\\
{\bfseries Return} $\tilde L$ and $ D$ (so that $\tilde L D \tilde L^T=A+J$) or $L=\tilde L \sqrt D$ (so that $L L^T=A+J$).\\
\end{tabbing}
The specific soft absolute value function $\text{smax}$ used here is given as
\[
\text{sabs}(x;u)= \frac{u}{\log(2)}\log\left(\exp\left(x\frac{\log(2)}{u}\right)+\exp\left(-x\frac{\log(2)}{u}\right) \right),
\]
Note that $\text{sabs}(x;u)\geq u$ with equality only for $x=0$. Moreover, $\text{sabs}(x;u)>|x| \;\forall\; |x|<\infty$.
\vspace{0.5cm}

\caption{\label{alg:modchol}Modified Cholesky decomposition. }
\end{algorithm*}

In the present paper, the elements of $\tilde{L}$ and $D$ are chosen
by the modified Cholesky algorithm according to the following
principles 
\begin{enumerate}
\item If it is known that the upper left sub-matrix $A_{1:K,1:K},$ for
some $1\leq K\leq d$, is positive definite, then $J_{j,j}=0$ for
$j=1,\dots,K$ and consequently\\ $(\tilde{L}D\tilde{L}^{T})_{1:K,1:K}=A_{1:K,1:K}$.
If no such information is available, then $K=0$.
\item The off-diagonal elements of $\tilde{L}D\tilde{L}^{T}$ are identical
to those of $A$.
\item If the finalized $D_{j,j}$, $j>K$ is found to be negative during
the modified Cholesky factorisation, which correspond to $A$
being negative- or indefinite, $D_{j,j}$ is substituted by a smooth
approximation to $\max(|D_{j,j}|,u_{j})$ where $u_{j}$ is a tuneable
lower bound. This approach is analogous to the flipping of signs of
negative eigenvalues used in the softmax metric of \citet{1212.4693}
\citep[see also][in the context of numerical optimization]{noce:wrig:1999},
but dissimilar in that $\tilde{L}$ is only a unit determinant transformation
whereas the corresponding transformation in the softmax metric is
orthonormal.
\end{enumerate}
\citet{gill_murray_1974,gilletal} included a further principle for
controlling numerical instabilities that can occur when $\{u_{j}\}_{j>K}$
are chosen to be close to machine precision \citep[see e.g.][page 108-109]{gilletal}.
In the present context, where $\{u_{j}\}_{j>K}$ are tuneable parameters
that typically take substantially higher values, these numerical instabilities
are not produced. A further rationale for not including the numerical stability-inducing
principle of \citet{gill_murray_1974,gilletal} is that it can introduce
artificially high values of $D_{j,j}$ in situations where the target
exhibits substantially different scales in different directions, and
consequently lead to slow exploration of the target. 

Moreover, \citet{gill_murray_1974,gilletal} do not include principle
1. This principle is, when applicable, very useful in the context
of RMHMC applied for hierarchical models such as LGMs where the posterior of the latent variables
($x_{1},\dots,x_{K}$), conditional on parameters ($x_{K+1},\dots,x_{d})$,
is log-concave, as unnecessary regularisation in the
form of lower bounds on $D_{j,j},\;j\leq K$ is not imposed. 

The modified Cholesky factorization is summarized in Algorithm \ref{alg:modchol}.
Steps 0-3 and 5 produce a standard $\bar{L}\bar{D}\bar{L}^{T}$, whereas
step 4 is specific to the modified Cholesky factorization. Assume first that $K>0$.  Then,
after $j\leq\min(K,d-1)$ iterations are completed, the sub-matrices
$\tilde{L}_{1:j,1:j}$ and $D_{1:j,1:j}$ are not written to in the
remaining iterations, and the intermediate values stored in $\tilde{L}_{j+1:d,1:d}$
and $D_{j+1:d,j+1:d}$ are not used for calculating $\tilde{L}_{1:j,1:j}$
and $D_{1:j,1:j}$. Thus, after iteration $j\leq\min(K,d-1)$, $\tilde{L}_{1:j,1:j}D_{1:j,1:j}(\tilde{L}_{1:j,1:j})^{T}=A_{1:j,1:j}$,
which is in agreement with principle 1. Further, principle 2 is fulfilled
as the diagonal elements of $A$ enter only in step 0,2 and the operations
on $D_{k,k},\;k>j$ in step 5 are strictly additive. Therefore, increasing
$D_{j,j}$ in step 4 effectively adds the difference between after
and before step 5 to the diagonal of the matrix being factorized,
i.e. this difference is identical to $J_{j,j}$ \citep{gilletal}.
Principle 3 follows directly from the application of the smooth absolute
value function $\text{sabs}$ in step 4, as $\text{sabs}(x;u)\geq u$
for $|x|<\infty$. 

In what follows, the metric tensor is taken to be of the form

\begin{multline}
G_{\mathbf{u}}(\mathbf{x})  =  L(\mathbf{x})L(\mathbf{x})^{T}=\tilde{L}(\mathbf{x})D(\mathbf{x})\tilde{L}(\mathbf{x})^{T}\\
=-\nabla_{\mathbf{x}}^{2}\log\tilde{\pi}(\mathbf{x})+J(\mathbf{x}),\label{eq:chol_def}
\end{multline}
and
\[
L(\mathbf{x})  =  \tilde{L}(\mathbf{x})\;\text{diag}\left(\sqrt{D(\mathbf{x})_{1,1}},\dots,\sqrt{D(\mathbf{x})_{d,d}}\right)
\]
where $\tilde{L}(\mathbf{x}),\;D(\mathbf{x})$ originate from applying
Algorithm \ref{alg:modchol} to $-\nabla_{\mathbf{x}}^{2}\log\tilde{\pi}(\mathbf{x})$
with regularisation parameter $\mathbf{u}$. Notice in particular that the log-determinant $\log|G_{\mathbf u}(\mathbf x)|$
required in (\ref{eq:hamiltonian}) is easily found as $\sum_{j=1}^d \log D(\mathbf x)_{j,j}$.
The resulting RMHMC method is referred to as
Modified Cholesky RMHMC (MCRMHMC).

\subsection{Low-dimensional illustration}

\begin{figure}
\centering{}\includegraphics[scale=0.45]{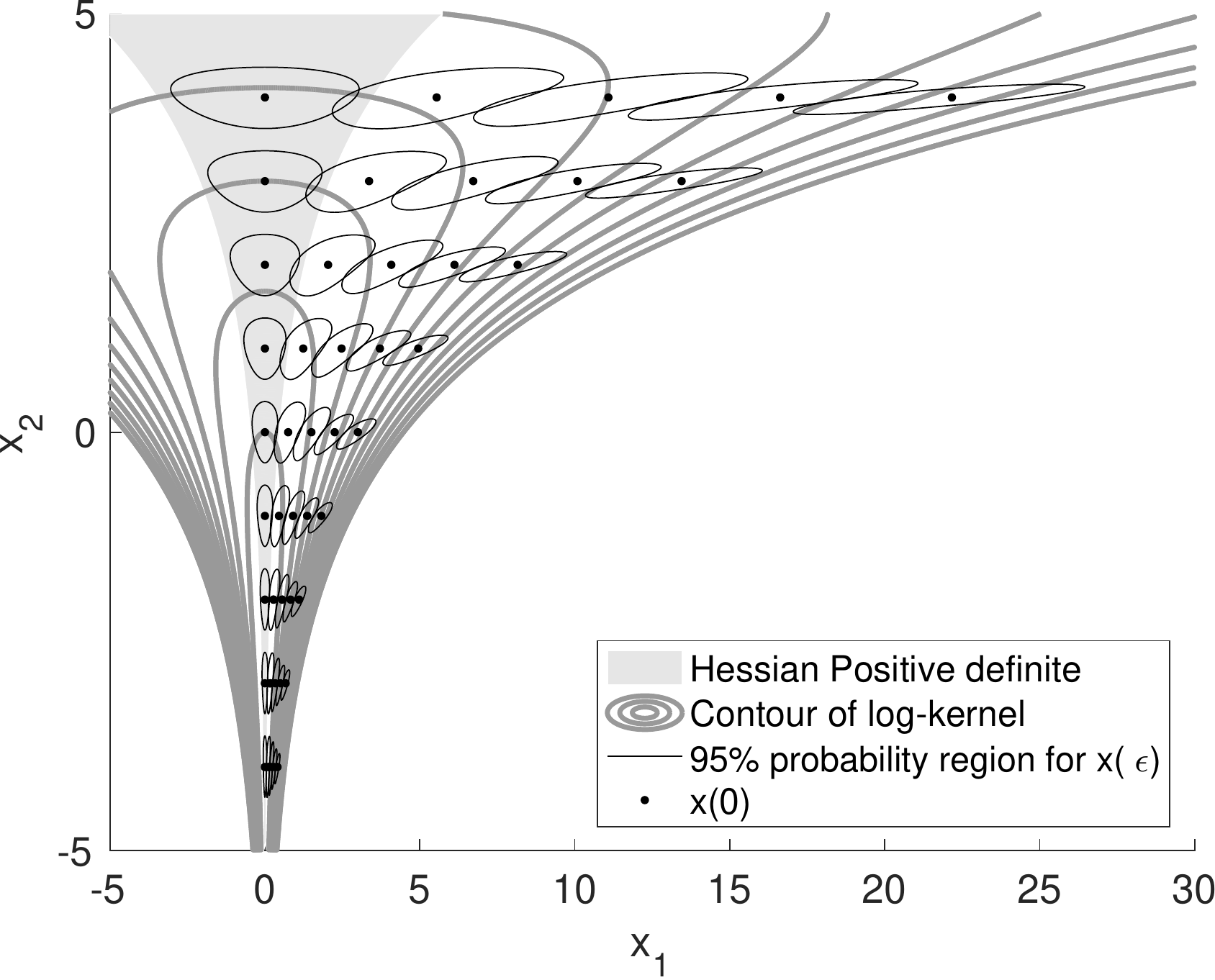}\caption{\label{fig:Illustration-funnel}Illustration of the effect of modified
Cholesky manifold scaling on RMHMC proposals for the bivariate funnel
distribution (\ref{eq:2d_funnel}). The MCRMHMC uses the tuning parameters
$K=1,u_{2}=1.0,$ $\varepsilon=0.15$. Only the right half-plane is considered
due to symmetry. Contours of the log-target (\ref{eq:2d_funnel})
are indicated by grey lines.  Dots indicate initial states $\mathbf{x}(0)$ and
closed black curves indicate corresponding 95\% probability regions
for $\hat{\mathbf{x}}(\varepsilon)|\mathbf{x}(0)$. The negative Hessian
$-\nabla_{\mathbf{x}}^{2}\log\tilde{\pi}(\mathbf{x})$ is positive
definite in the region shaded with light gray. Note that corresponding 95\% probability region
curves for EHMC methods (such as the Stan NUTS) would have the same (elliptical) shape for all initial states.}
\end{figure}
To illustrate the proposed methodology, consider a bivariate
version of the \citet{neal2003} funnel distribution, given as
\begin{equation}
\log\tilde{\pi}(\mathbf{x})=-\frac{x_{1}^{2}}{2\exp(x_{2})}-\frac{x_{2}}{2}-\frac{x_{2}^{2}}{18}.\label{eq:2d_funnel}
\end{equation}
Namely, $x_{1}|x_{2}\sim N(0,\exp(x_{2}))$, $x_{2}\sim N(0,3^{2})$.
This model displays substantially different scales in $x_{1}$ depending
on the value of $x_{2}$, and therefore illustrates how joint sampling
of variables along with the variance parameter of these variables (e.g. latent field
and the variance of the latent field) poses substantial
problems for MCMC methods that do not adapt to local scaling properties.

Methods based on mode and (negative inverse) Hessian at the mode \citep[Chapter 13.3]{GelmanBDA3}
for scaling MCMC proposals are not reliable for such targets. I.e.
the target has mode at $\left[0,-4.5\right]^{T}$ whereas $E(\mathbf{x})=\left[0,0\right]^{T}$.
The negative inverse Hessian at the mode variance approximation yields $\text{diag}(0.0111,9)$
whereas $Var(\mathbf{x})=\text{diag}(90.01,9).$ It is seen that the
mode is shifted in the $x_{2}$-direction toward the smaller scale
region (negative $x_{2}$) as smaller scales yield higher values
of $p(x_{1}|x_{2})$. The variance in $x_{1}$-direction indicated
by the inverse negative Hessian at the mode is also very different
from $Var(x_{1})$, but one can argue that none of these variances
are very informative for globally scaling MCMC proposals due to the
different scales in $x_{1}$-direction determined by $x_{2}$.

Since $x_{1}|x_{2}$ is Gaussian, it is clear that\\ $(-\nabla_{\mathbf{x}}^{2}\log\tilde{\pi}(\mathbf{x}))_{1,1}>0$,
and therefore $K=1$ is used to implement MCRMHMC for this model. Figure
\ref{fig:Illustration-funnel} shows 95\% probability regions for
single (time integration) step proposals of MCRMHMC, for a selection
of initial configurations (indicated by dots) for target (\ref{eq:2d_funnel}).
It is seen that the metric tensor of MCRMHMC appropriately scales the
proposals and aligns the proposal distributions to the local geometry
of the target. The negative Hessian is positive definite only in region
shaded with light gray ($x_{1}\in(\mp(\sqrt{2}/3)\exp(x_{2}/2))$),
and it is seen that the negative Hessian in conjunction with the modified
Cholesky factorization also produces useful scale information when
the Hessian is indefinite. This latter observation is very much in
line with the numerical optimization literature
on modified Newton methods \citep[Chapter 6]{noce:wrig:1999}. Moreover,
due to the relatively high degree of regularisation in the $x_{2}$-direction
($u_{2}=1.0$), no problems related to close to
zero eigenvalues near boundaries of the shaded region are seen \citep{SJOS:SJOS12204}. 

\subsection{Discussion of the modified Cholesky factorization}\label{sec:chol_discussion}

Several issues related to the modified Cholesky factorization in Algorithm
\ref{alg:modchol} and its application in the RMHMC context require
further clarification at this point. First, it is clear that Algorithm
\ref{alg:modchol} is \emph{not invariant to reordering of the variables
in $\mathbf{x}$,} a property that is often imposed in optimisation contexts
 via symmetric
row and column interchanges \citep{gilletal,noce:wrig:1999}.
Such row and column interchanges are not
applied here as they 1) introduce discontinuities
in $\nabla_{\mathbf{x}}H$ which in turn makes simulating the flow
associated with (\ref{eq:x_evol},\ref{eq:p_evol}) more difficult,
2) disturb any exploitation of $(K\times K$) positive definite upper
left sub-matrices, 3) make exploitation of sparsity of the negative
Hessian substantially less effective as the sparsity pattern of the
modified Cholesky factorizations changes. Moreover, a further rationale
for including such symmetric row- and column interchanges is to avoid
numerical instabilities associated with small finalized $D_{j,j}$s,
but in the present context such problems can be tuned away by appropriate
increases in $u_{j}$. 
\begin{figure}
\begin{centering}
\includegraphics[scale=0.5]{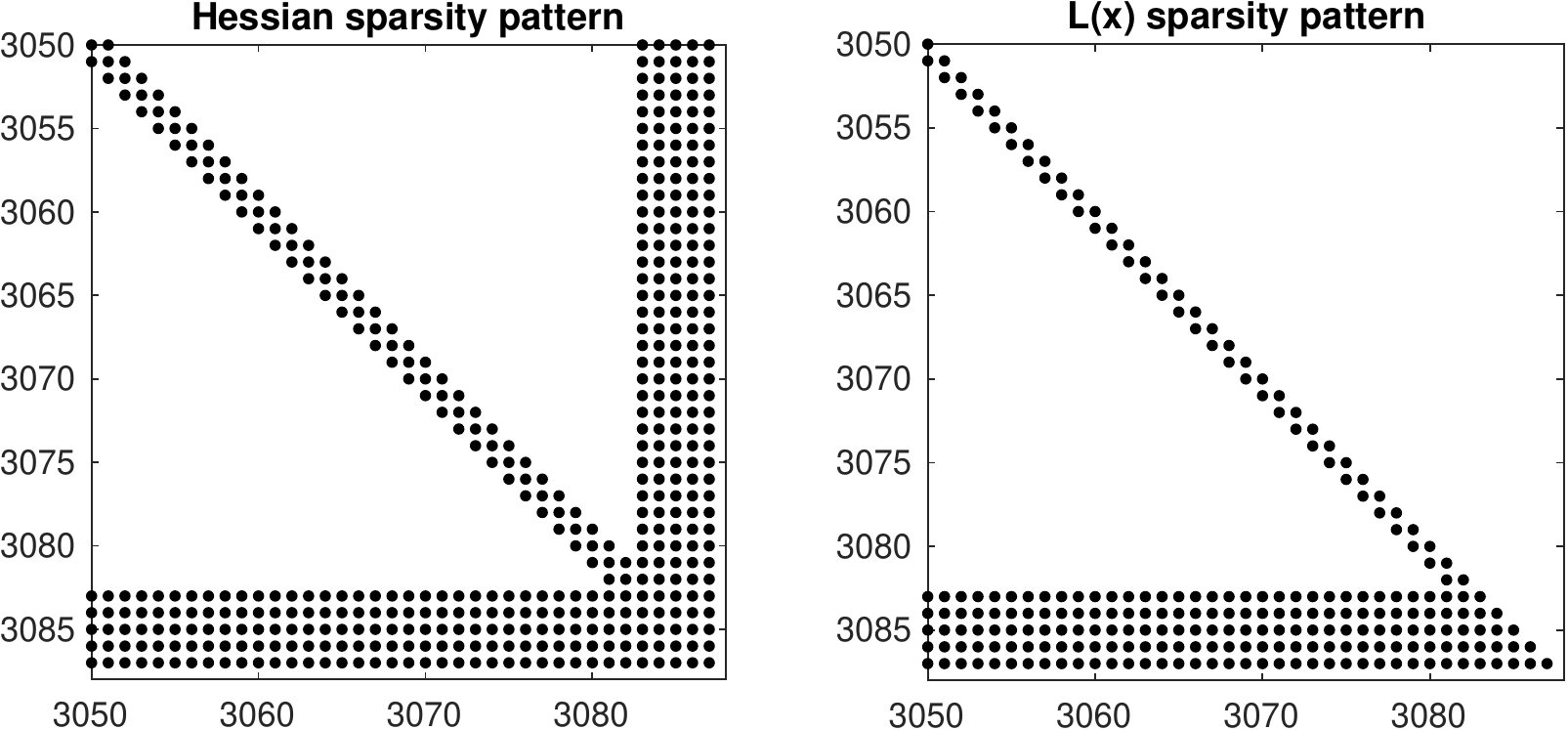}\caption{\label{fig:Sparsity-patterns-for}Sparsity patterns for the log-target
Hessian and its associated modified Cholesky factor $L(\mathbf{x})$
for the model considered in Section \ref{sec:Application-to-a}. Non-zero
elements are indicated by dots. In order to improve readability, only the 
lower right (3050:3087,3050:3087) submatix is shown in both cases. The
patterns repeats along the borders and (sub,sup)-diagonals in the complete
matrices. For the Hessian, only 0.4\% of the elements are non-zero, whereas only
0.5\% of the lower triangular elements of $L(\mathbf x)$ are non-zero.}
\par\end{centering}
\end{figure}

Given that Algorithm \ref{alg:modchol} is not invariant to reordering
of variables, it reasonable to ask what is an appropriate ordering
of the variables? The short answer is that this is a tradeoff between
exploiting any possible positive definite sub-matrices by putting
the associated variables first in $\mathbf{x}$, and using orderings
of the variables so that $L(\mathbf{x})$ is as sparse as possible
\citep{davis_sparse}. Fortunately, in the context of hierarchical
models and especially for LGMs, the means to these two objectives often align well, in the
sense that the latent vector, conditional on parameters, often
is both associated with a positive definite negative Hessian and \citep[possibly
after an appropriate internal reordering using e.g. the AMD algorithm of][Chapter 7]{davis_sparse}
has a Cholesky factorization with exploitable sparsity structure 
\citep[see][where both of these properties are exploited in the LGM context in the
Laplace approximation used for calculating marginal posteriors of the parameters]{RSSB:RSSB700}.
Thus, assuming the above situation, a rule of thumb will be to put
the latent variables first in $\mathbf{x}$ and let the last elements
of $\mathbf{x}$ be the parameters. This strategy leads to the rows
of $L(\mathbf{x})$ corresponding to the latent variables being sparse,
and the remaining (typically few) rows corresponding to parameters
being dense. To illustrate this rule of thumb procedure, Figure \ref{fig:Sparsity-patterns-for}
displays the sparsity patterns of the log-target negative Hessian
and associated Cholesky factor $L(\mathbf{x})$ for the non-linear
state space model discussed in detail in Section \ref{sec:Application-to-a}.
The model consists of $3082$ latent variables with first order (time
series) Markov structure (put first in $\mathbf{x}$), and
5 parameters (put last in $\mathbf{x}$). Due to the Markovian
structure, the sub-matrix corresponding to the latent variables is
tri-diagonal, and consequently the Cholesky factor has only non-zero
elements on the diagonal and the first sub-diagonal. This property
is retained for the Cholesky factor of the complete negative Hessian,
where only the rows corresponding to the parameters are dense.

A discussion of the vector of regularisation parameters $\mathbf{u}$
is also in order. Recall that Algorithm \ref{alg:modchol} requires
$d-K$ regularisation parameters, where the $\left\{ u_{j}\right\} _{j=K+1}^{d}$
should reflect the potentially very different scaling of $\left\{ x_{j}\right\} _{j=K+1}^{d}$.
Including a potentially large number of regularisation parameters
naturally comes both with added flexibility and potential for very
high-fidelity sampling when tuned properly. On the other hand,
including many regularisation parameters may lead to time-consuming tuning efforts, in particular
since the interpretation of $\left\{ u_{j}\right\} _{j=K+1}^{d}$
is not as straight forward as in the softmax metric based on full
spectral decompositions \citep{1212.4693}. Though, of course, all
active regularisation parameters can be set equal, and thus reduce
to a single regularisation parameter as is the case in the softmax
metric of \citet{1212.4693}, in many cases this may lead to suboptimal
sampling in a RMHMC context as unnecessary regularisation will lead
to slower and more oscillating exploration of the target. 

To further
illustrate the advantages of this regularisation scheme,
consider the toy model discussed in detail in Section \ref{subsec:Funnel-AR(1)-model},
consisting of a Gaussian zero mean AR(1) model with autocorrelation
0.999 as $\mathbf{x}_{1:d-1}$, and let $x_{d}$ be the logarithm
of the innovation precision of said AR(1) model. For $d=1000$, $x_{d}=-3.0$
and $\mathbf{x}_{1:d-1}$ simulated from the true model, 
the eigenvalues of the negative Hessian
are (in ascending order) $(-0.067,\;1.35e-7,\;7.22e-7,\dots)$, whereas
the eigenvalues of the negative Hessian associated with $\mathbf{x}_{1:d-1}$ are
$(1.35e-7,\;7.22e-7,\dots)$. Given that $\mathbf{x}_{1:d-1}$
is jointly Gaussian, the negative Hessian of $\mathbf{x}_{1:d-1}$
is by construction positive definite, but with smallest eigenvalue
several orders of magnitude smaller than the absolute value of the single negative
eigenvalue associated with $\mathbf{x}$. The modified Cholesky approach,
with $K=d-1$, will exactly reproduce the precision matrix of $\mathbf{x}_{1:d-1}$
in $G_{1:d-1,1:d-1}$ and only apply regularisation in the dimension
corresponding to the log-precision parameter. Regularisation in eigen-space,
on the other hand, is likely to disturb the representation of the precision
matrix of $\mathbf{x}_{1:d-1}$ in $G$ substantially, as it is likely
that regularisation at least a few orders of magnitude smaller than
the negative eigenvalue needs to be applied. The disturbed representation of
these eigenvalues will lead to less efficient sampling as the resulting
RMHMC algorithm will to a lesser degree align the Hamiltonian dynamics
with the strong dependence among $\mathbf{x}_{1:d-1}$. 

\subsection{Implementation and tuning\label{subsec:Implementation-and-tuning}}

The prototype large scale code used in the reminder of this paper
is implemented in C++ and uses a sparsity-exploiting implementation
of Algorithm \ref{alg:modchol} that is derived from the function
\texttt{cs\_chol} of the \texttt{csparse} library \citep{davis_sparse}.
Since the sparsity pattern of $L$ is identical to that of a conventional
sparse Cholesky factorization, the functions of the \texttt{csparse}
library for calculating sparsity patterns, speeding up computations
(elimination trees) and other numerical tasks such as triangular solves
can be used directly. 

The code makes use of the template capabilities of the C++ language
so that a single code is maintained both for \texttt{double} numerics for calculating
$H$ and $\nabla_{\mathbf{p}}H$, and with automatic differentiation
(AD) types for calculating $\nabla_{\mathbf{x}}H$. Specifically,
the AD tool \texttt{adept} \citep{Hogan:2014:FRA:2639949.2560359} is used for the latter task. Finally,
in the current version, $-\nabla_{\mathbf{x}}^{2}\log\tilde{\pi}(\mathbf{x})$
is hand-coded. However, in future versions, this task will also be
automated using AD-software capable of exploiting the sparsity
of the Hessian \citep{grie:2000}. 

As discussed above, the methodology involves a number of tuning parameters,
which need to be adjusted for each particular target distribution.
Here, a method for such adjustment during a warm up phase is provided,
which has been used for the computations described in the reminder
of the paper. Firstly, the ordering of variables and selection of
$K$ is done mainly based on insight into the problem at hand, with
$K$ latent variables typically taken first in $\mathbf{x}$, and
parameters, in particular scales and correlations, put last in $\mathbf{x}$.
Subsequently, too high initial guesses of $K$ can be reduced to $j-1$
during warm up if $D_{j,j}<0,$$j\leq K$ at step 4 of Algorithm for
some $\mathbf{x}$. If such a situation is encountered during the
post-warm up phase, the MCMC simulation needs to be restarted with
reduced $K$.

Secondly, the joint tuning of $\varepsilon$, $l$ and $\left\{ u_{j}\right\} _{j=K+1}^{d}$
requires some more attention. The heuristic strategy adopted here
is based on the following three steps:
\begin{itemize}
\item \textbf{Select some reference $\varepsilon$ and $l$}, e.g. $\varepsilon=0.5d^{-0.25}$
and $E(l)=\lfloor1.5/\varepsilon\rfloor$. The expressions
for $\varepsilon$, $l$ are informed firstly by the fact that for any Gaussian
target, say $N(\mu,\Sigma)$, then $G(\mathbf{x})=\Sigma^{-1}$ and
the integrator (\ref{eq:p_update1}-\ref{eq:GL_explicit}) reduces
to the conventional leap frog integrator for separable Hamiltonian
systems. In this situation, it is known that an asymptotic expression
for the expected acceptance probabilities for proposals independent
of current configuration ($\varepsilon l\approx\pi/2)$ is $2-2\Phi(\varepsilon^{2}\sqrt{d}/8)$
\citep[see e.g.][]{beskos2013,doi:10.1080/03610918.2017.1283703}, which when solved for
$\varepsilon$ with acceptance probability 0.95 yields $\varepsilon\approx0.7d^{-0.25}$.
In practice, it is usually the case that slightly smaller values of $\varepsilon$
are needed for models with non-constant curvature and hence $\varepsilon=0.5d^{-0.25}$
is taken as a reasonable step size. 
Secondly, the reference expression for the number of integration steps is
informed by the $l\varepsilon\approx\pi/2$ needed in the
Gaussian case, whereas the connection to Newton's method in optimization
indicates $l\varepsilon\approx\sqrt{2}$. 
\item \textbf{Tune $\left\{ u_{j}\right\} _{j=K+1}^{d}$ using the reference
values of $\varepsilon$ and $l$}, with the objective of taking $\left\{ u_{j}\right\} _{j=K+1}^{d}$
as small as possible while ensuring the Hamiltonian is well-behaved
enough for the fixed point iterations associated with (\ref{eq:implicit_1}-\ref{eq:implicit_2})
to converge. In practice, this tuning is implemented by initially
setting each active $u_{j}$ to a small number, say $\exp(-20)$,
to avoid unnecessary and computationally wasteful regularisation,
and run a number of warm up iterations while increasing the relevant
$u_{j}$ by a factor say $\exp(1)$ whenever a divergence in the fixed
point iterations is encountered. Divergences in the fixed point iterations
for both (\ref{eq:implicit_1}) and (\ref{eq:implicit_2}) indicate
that $\Vert J(\mathbf{x},\mathbf{p})\Vert>1$ where $J(\mathbf{x},\mathbf{p})=\frac{\varepsilon}{2}\nabla_{\mathbf{x},\mathbf{p}}^{2}H(\mathbf{x},\mathbf{p})=\frac{\varepsilon}{2}\nabla_{\mathbf{x}}[G^{-1}(\mathbf{x})\mathbf{p}]$
around the sought fixed point. Moreover, for small $|D_{j,j}|/u_{j}$
(relative to say 10) and small $u_{j}$, $\frac{d}{dz}[\text{sabs}(z;u_{j})]^{-1}|_{z=D_{j,j}}$
contributes substantially to $\Vert J(\mathbf{x},\mathbf{p})\Vert$
and an ideal approach for choosing which $u_{j}$ to increase would
be to measure the sensitivity of $\Vert J(\mathbf{x},\mathbf{p})\Vert$
to each of the active $u_{j}$s. Rather than computing the norm of
the complete Jacobian, a heuristic approach is taken, which involves increasing $u_{\hat{j}}$
for\\ $\hat{j}=\arg\max_{K<j\leq d}|\frac{d}{dz}[\text{sabs}(z;u_{j})]^{-1}|_{z=D_{j,j}}|$
whenever a divergence is detected. It is worth noticing that, in failing
to take into account the sensitivity of the negative Hessian on $\mathbf{x}$
and the fact that $D_{j,j}$ is also used in subsequent iterations
in Algorithm \ref{alg:modchol}, this approach is likely to be suboptimal
with respect to selecting the relevant $u_{j}$. This is in particular
the case when the target under consideration contains strong dependencies,
and thus basing such a selection mechanism on a more direct approximation
to $\Vert J(\mathbf{x},\mathbf{p})\Vert$ is scope for future research.
\item \textbf{Tune $\varepsilon$ and $l$ for fixed $\left\{ u_{j}\right\} _{j=K+1}^{d}$},
with the objective of controlling the acceptance probability and producing
low-autocorrelation samples using standard procedures. In particular,
this part of the tuning is readily automated using dynamic selection
of $l$ \citep{1304.1920,JMLR:v15:hoffman14a,1601.00225} and tuning
$\varepsilon$ toward a given acceptance probability e.g. using the
dual averaging algorithm of \citet{JMLR:v15:hoffman14a}.
\end{itemize}

\section{Simulation study\label{sec:Simulation-study}}

To benchmark the proposed methodology against common, general purpose
MCMC methods, two challenging toy models are considered. These exhibit, 
in the first case, strong non-linear dependence, and in the second
case, a ``funnel'' effect. The contending methods are chosen as
they are routinely used in diverse high-dimensional applications. Throughout 
this section, all methods except Stan
were implemented in C++ and compiled with the same compiler, and with
the same compiler settings. The Stan computations were done using the
R-interface \texttt{rstan}, version 2.15.1, running under \texttt{R} version 
3.3.3. The computer used for all computations
in this section was a 2014 Macbook Pro with a 2.0 GHz Intel Core i7
processor.

\subsection{Twisted Gaussian mean AR(1) model\label{subsec:Twisted-Gaussian-mean}}
\begin{table*}
\centering{}%
\begin{tabular}{lccccccccccc}
\hline 
Method & \# MCMC & \multicolumn{2}{c}{CPU time}  & \multicolumn{2}{c}{$\underset{1\leq i\leq d-1}{\min}ESS(x_{i})$} & 
\multicolumn{2}{c}{$ESS(x_{d})$} & \multicolumn{2}{c}{$\frac{\underset{1\leq i\leq d-1}{\min}ESS(x_{i})}{\text{CPU time}}$} & \multicolumn{2}{c}{$\frac{ESS(x_{d})}{\text{CPU time}}$}\tabularnewline
 &  iterations &\multicolumn{2}{c}{(seconds)} & &  &  &  & \tabularnewline
Across replica & & min & mean & min & mean& min & mean& min & mean& min & mean\\
\hline 
\multicolumn{12}{c}{$d=10$}\tabularnewline
\hline 
MCRMHMC  & 1000 & 3.3 & 3.4 & 603 & 813 & 891 & 981 & 177 & 239 & 259 & 289\tabularnewline
Gibbs  &  5000000 & 2.3 & 2.3 & 33 & 94 & 59 & 146 & 14 & 40 & 25 & 63\tabularnewline
EHMC  & 5000 & 4.7 & 4.8 & 1056 & 1182 & 630 & 700 & 220 & 246 & 132 & 146 \tabularnewline
Stan & 5000 & 1.4 & 1.4 & (4312) & (4800) & (4465) & (4833) & (3137) & (3355) & (3144) & (3379)\\
EigenRMHMC & 1000 & 425 & 432 & 59 & 93 & 59 & 96 & 0.14 & 0.21 & 0.14 & 0.22\\
\hline 
\multicolumn{12}{c}{$d=100$}\tabularnewline
\hline 
MCRMHMC  & 1000 & 65 &   66 &  756 &  873 &  843 &  954 &   11 &   13 &   13 & 14\tabularnewline
Gibbs  &  5000000 & 22 &  22  &  (7.2) & (24) & (9.6) & (26) & (0.32) & (1.1) & (0.43) & (1.2) \tabularnewline
EHMC  & 5000 &  75 &   76 &  460 &  671 &  570 &  775 &    6.0  &   8.8  &   7.6  &  10\tabularnewline
Stan & 5000 & 12 &14  & (4648) & (4868) & (4531) & (4870) & (315) & (349) & (324) & (349) \\
\hline 
\multicolumn{12}{c}{$d=1000$}\tabularnewline
\hline 
MCRMHMC  &     1000 &  1435 &1456 &451 &723 &728 & 879 & 0.31 & 0.50 & 0.50 & 0.60 \tabularnewline
Gibbs  & 5000000 & 223 & 224 & (4.0) & (15) &  (3.9) & (12) & (0.02) & (0.07) & (0.02) & (0.06) \tabularnewline
EHMC  & 5000 & 1855& 1897& 46 & 214 & 90 & 268 & 0.02 & 0.11 & 0.05 & 0.14   \tabularnewline
Stan & 5000 & 225 & 255 & (4529) & (4786) & (4607) & (4852) & (17) & (19) & (17) & (19)\\
\hline 
\end{tabular}\caption{\label{tab:Results-for-the-twisted}Results for the twisted Gaussian
mean AR(1) model experiment. Each combination of method and $d\in\{10,100,1000\}$
was repeated 10 independent times and the numbers presented are minimum (min) and means (mean)
across these replica. Visual inspection
of output (Figures \ref{fig:Traceplots-(left),-sample-twisted1} -
\ref{fig:Traceplots-(left),-sample-twisted3}) and Kolmogorov-Smirnov tests 
(applied for MCRMHMC and Stan)
indicate that some samplers failed to explore the relevant support, and the corresponding
numbers are given in parentheses. The results for the Gibbs
sampler are only recorded every 1000th iteration, thus the maximum
ESSs for Gibbs are 5000. Tuning parameters for MCRMHMC are: $d=10$: $u_{n}=\exp(3.5),\varepsilon=0.4,l\sim U(20,30)$, $d=100$: 
$u_{n}=\exp(3.5),\varepsilon=0.15,l\sim U(60,80)$, 
$d=1000$: $u_{n}=\exp(3.5),\varepsilon=0.1,l\sim U(130,160)$. For Gibbs,
tuning parameters are: $d=10$: $r_{Gibbs}=0.65$, $d=100$: $r_{Gibbs}=0.4$ and $d=1000$: $r_{Gibbs}=0.05$. 
For EHMC, tuning parameters are: $d=10$: $\varepsilon=0.02,l\sim U(700,1000)$, $d=100$: $\varepsilon=0.01,l\sim U(1500,2000)$, and $d=1000$: 
$\varepsilon=0.007,l\sim U(4000,6000)$. Finally for EigenRMHMC, $d=10$: $u=\exp(3)$, $\varepsilon=0.3$ and $l\sim U(50,100)$.}
\end{table*}
\begin{figure*}
\centering{}\includegraphics[scale=0.7]{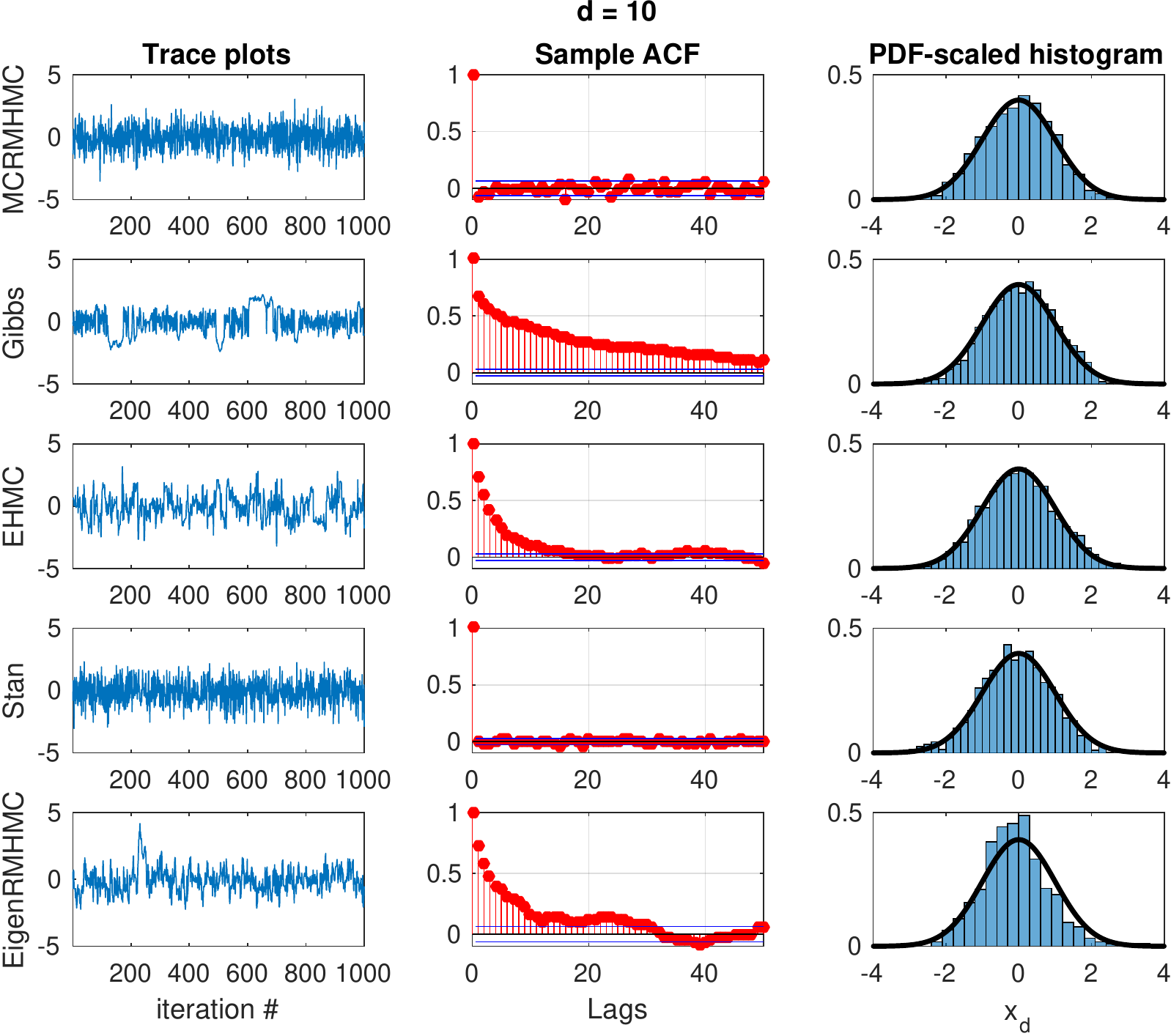}\caption{\label{fig:Traceplots-(left),-sample-twisted1}Traceplots (left), sample
autocorrelation functions (center) and pdf-scaled histograms (right)
of MCMC iterations for $x_{d}$ for $d=10$. Under target characterized by (\ref{eq:twisted-ar-1}-\ref{eq:twisted-ar-3}),
$x_{d}$ has a standard Gaussian marginal distribution, which is indicated
as a reference for the histograms. }
\end{figure*}
\begin{figure*}
\centering{}\includegraphics[scale=0.7]{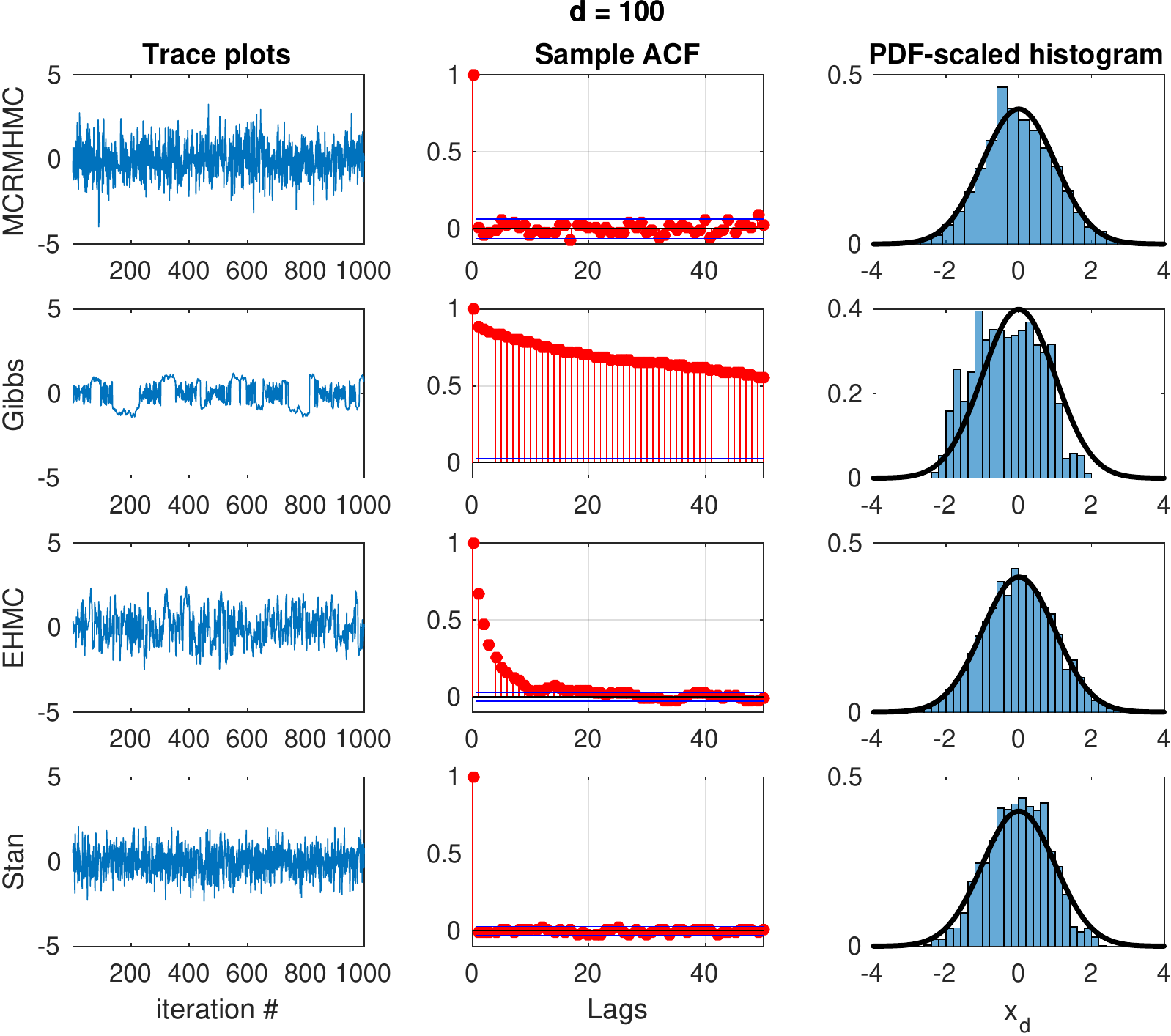}\caption{\label{fig:Traceplots-(left),-sample-twisted2}Traceplots (left), sample
autocorrelation functions (center) and pdf-scaled histograms (right)
of MCMC iterations for $x_{d}$ for $d=100$. Under target characterized by (\ref{eq:twisted-ar-1}-\ref{eq:twisted-ar-3}),
$x_{d}$ has a standard Gaussian marginal distribution, which is indicated
as a reference for the histograms. }
\end{figure*}
\begin{figure*}
\centering{}\includegraphics[scale=0.7]{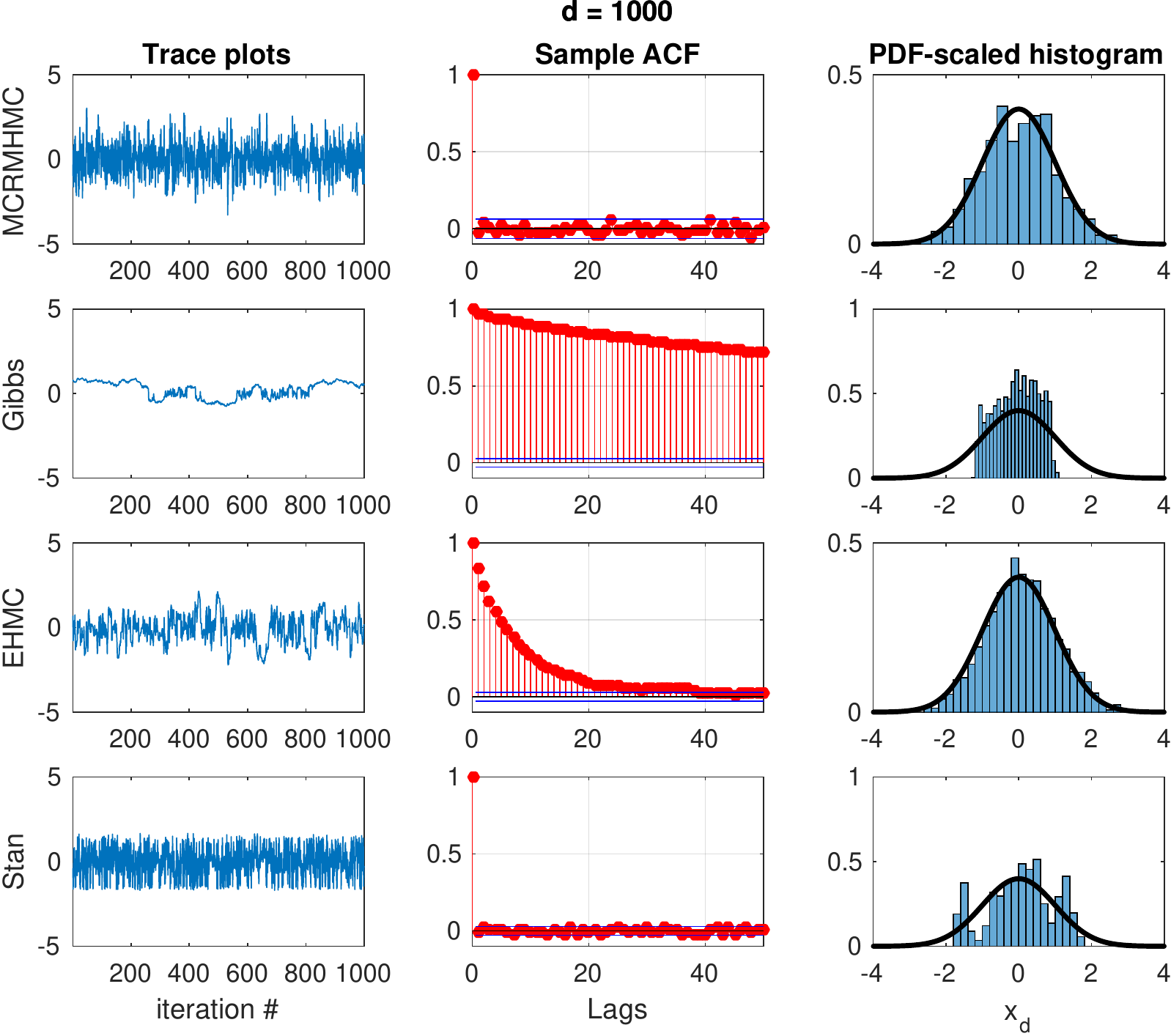}\caption{\label{fig:Traceplots-(left),-sample-twisted3}Traceplots (left), sample
autocorrelation functions (center) and pdf-scaled histograms (right)
of MCMC iterations for $x_{d}$ for $d=1000$. Under target characterized by (\ref{eq:twisted-ar-1}-\ref{eq:twisted-ar-3}),
$x_{d}$ has a standard Gaussian marginal distribution, which is indicated
as a reference for the histograms. }
\end{figure*}
This first toy model is given by
\begin{multline}
x_{i}|x_{i-1},x_{d} \\
\sim N\left((x_{d}^{2}-1)+0.95(x_{i-1}-(x_{d}^{2}-1)),\frac{1-0.95^{2}}{100}\right),\\
\;i=2,\dots,d-1,\label{eq:twisted-ar-1}
\end{multline}
\begin{align}
x_{1}|x_{d} & \sim N\left(x_{d}^{2}-1,\frac{1}{100}\right),\label{eq:twisted-ar-2}\\
x_{d} & \sim N(0,1).\label{eq:twisted-ar-3}
\end{align}
I.e. conditionally on $x_{d}$, $\mathbf{x}_{1:d-1}$ is a Gaussian
AR(1) process with autocorrelation 0.95, marginal standard deviation
0.1 and mean $x_{d}^{2}-1$. Similar specifications, but without autocorrelation,
have previously been considered as test cases
for MCMC algorithms \citep{Haario1999,1304.1920}. The model may be considered as a very 
simple hierarchical model,
where $\mathbf x_{1:d-1}$ are latent variables, $x_d$ is a parameter, and no observations are modelled
(in order to retain analytical marginals).

The contending methods are chosen as they share scope with respect to generality. Specifically,
Gibbs sampling, 
EHMC with identity mass matrix, Stan and RMHMC using full spectral decomposition (EigenRMHMC)
similar to \cite{1212.4693}, in addition to the proposed methodology were considered. 
The Gibbs sampler (Gibbs) was implemented
using single dimension updates in order to mimic a situation where
$\mathbf{x}_{1:d-1}$ are latent variables with intractable joint conditional
posterior. Specifically, for updating $x_{i}|\mathbf{x}_{-i},\;i=1,\dots,d-1$,
exact Gaussian updates were used. 
For updating $x_{d}|\mathbf{x}_{1:d-1}$,
a Gaussian random walk Metropolis Hastings method, with proposal standard
deviation $r_{Gibbs}$ was used. Here, $r_{Gibbs}$ was tuned to
produce update rates of 20-30\%. Due to the slow mixing, but low per
iteration computational cost, only every 1000th Gibbs iteration was recorded.

For EHMC, uniformly distributed numbers of time integration
steps $l$ were used, where the notation $U(a,b)$ denotes uniform on the
integers between $a$ and $b$ including $a,b$. In addition, the
integration step sizes were jittered with $\pm15\%$ uniform noise
\citep{1206.1901}. The distributions for $l$ and $\varepsilon$
were tuned to target an acceptance rate around 60\% and
to produce MCMC samples that were well separated in the $x_{n}$-direction.

Stan was run using 10000 iterations, where the first 5000 (warm-up) iterations
were used for tuning the sampler. The \texttt{diag\_e} metric (i.e. general
fixed diagonal scaling matrix) and otherwise default settings were used. 
The warm-up iterations are not included in the reported CPU times. 

The negative Hessian of the log-density
of $\mathbf{x}_{1:d-1}|x_{d}$ is positive definite, which enables
$K=d-1$ for\\ MCRMHMC. The sparsity patterns of the negative Hessian and modified
Cholesky factor $L$ are similar to those of Figure \ref{fig:Sparsity-patterns-for}
with the exception that this model only has one ``parameter'' with
corresponding dense $d$th row in Hessian and $L$. The values for
$u_{d}$ were found by successively increasing $u_{d}$ until the
iterative solvers in the time integrator no longer diverged for a
reference value of $\varepsilon$. Also here, uniformly distributed
numbers of time integration steps and $15\%$ jittered step sizes were used.
The integration step size was taken to produce acceptance rates of
around 95\% and the tuning of the distribution of $l$ was again taken
to produce high fidelity samples for $x_{d}$. 

EigenRMHMC was implemented using full spectral decomposition, followed by
applying the sabs-function (see Algorithm 1) with parameter $u$ to each
eigenvalue to produce a positive definite metric tensor. Other than the
metric tensor, MCRMHMC and EigenRMHMC are identical. In particular
$u$ was tuned by increasing $u$ until the iterative solvers no longer diverged
and the integration step size was tuned toward acceptance rates around 95\%.
For both MCRMHMC and EigenRMHMC, AD was applied to the complete code (i.e. including
modified Cholesky factorization- and spectral factorization codes). 

The experiment was run for each of $d\in\{10,100,1000\}$. EigenRMHMC
was only considered in the $d=10$ case, as very long computing times 
were required in higher dimensions.
In all considered cases, each method was run 10 independent times. For each
method except Stan, realizations
from the target distribution were used as initial values for the MCMC methods,
and therefore no warm up iterations were performed. To compare the
performances of the MCMC methods, the effective sample size (ESS),
calculated using the monotone sample autocorrelation estimator of
\citet{geyer1992} was employed. 

The results are presented in Table \ref{tab:Results-for-the-twisted},
and trace plots, sample autocorrelation functions and histograms for
$x_{d}$ are presented in Figures \ref{fig:Traceplots-(left),-sample-twisted1} -
\ref{fig:Traceplots-(left),-sample-twisted3}.
First, it is seen from Figures \ref{fig:Traceplots-(left),-sample-twisted1} -
\ref{fig:Traceplots-(left),-sample-twisted3}
that the Gibbs sampler, even after 5 million iterations, fail to properly
explore the target distribution. Also Stan seem to produce samples that do not
fully explore the target distribution, which is seen most clearly from Figure 
\ref{fig:Traceplots-(left),-sample-twisted3}.
Due to the close to zero autocorrelation of
the Stan and MCRMHMC MCMC samples, Kolmogorv-Smirnov tests with null hypothesis
being that samples representing $x_d$ are standard normal were conducted. The null hypothesis
is strongly rejected for Stan in all cases, whereas not rejected for MCRMHMC in all cases. 
Thus, only MCRMHMC, EHMC and EigenRMHMC appear to produce
accurate representations of the target distribution, with MCRMHMC producing
uniformly highest ESSes and ESSes per computing time for $d=100,1000$,
both in minimum and mean across replica.
Moreover, the relative ESS per CPU time for MCRMHMC and EHMC appears
to increase with increasing dimension, with MCRMHMC being a factor
4 faster for $d=1000$. Further, visual inspection of the
Hamiltonian dynamics trajectories (unreported) 
shows a very coherent and fast exploration
of the target by the MCRMHMC, whereas the trajectories of EHMC and EigenRMHMC are
oscillating due to the strong dependency structure, and therefore
lead to a slow exploration of the target. In the case of EigenRMHMC, it seems the regularisation
parameter needed to make the fixed point iterations of integrator converge also strongly
influences the representation of the AR(1) model precision in the implied $G(\mathbf x)$,
which slows down the exploration.

\subsection{Funnel AR(1) model\label{subsec:Funnel-AR(1)-model}}

The second toy model considered is a zero mean Gaussian AR(1) model with
autocorrelation $0.999$ jointly with the innovation precision parameter,
where the latter has a gamma(1,0.1) prior: 
\begin{multline}
x_{i}|x_{i-1},x_{d} \sim N\left(0.999x_{i-1},\frac{1}{\exp(x_{d})}\right),\\ \;i=2,\dots,d-1,\label{eq:funnel_def_1}
\end{multline}
\begin{align}
x_{1}| x_{d}&\sim N\left(0,\frac{1}{\exp(x_{d})(1-0.999^{2})}\right),\label{eq:funnel_def_2}\\
\exp( x_{d})&\sim\text{gamma}(1,0.1).\label{eq:funnel_def_3}
\end{align}
This model exhibit a ``funnel'' nature \citep{neal2003,1304.1920}
as the marginal standard deviation of $x_{i},\;i=1,\dots,d-1$ varies
by two orders of magnitude between the $0.001$ and $0.999$ quantiles
of $x_{d}$. Moreover, there is a very strong dependence among $x_{i},\;i=1,\dots,d-1$
which adds further complications for many MCMC methods. Thus, also
this target distribution shares many of the features of the (joint
latent variables and parameters) target distribution associated with
hierarchical models. 
\begin{table*}
\centering{}
\begin{tabular}{lccccccccccc}
\hline 
Method & \# MCMC & \multicolumn{2}{c}{CPU time}  & \multicolumn{2}{c}{$\underset{1\leq i\leq d-1}{\min}ESS(x_{i})$} & 
\multicolumn{2}{c}{$ESS(x_{d})$} & \multicolumn{2}{c}{$\frac{\underset{1\leq i\leq d-1}{\min}ESS(x_{i})}{\text{CPU time}}$} & \multicolumn{2}{c}{$\frac{ESS(x_{d})}{\text{CPU time}}$}\tabularnewline
 &  iterations &\multicolumn{2}{c}{(seconds)} & &  &  &  & \tabularnewline
Across replica & & min & mean & min & mean& min & mean& min & mean& min & mean\\
\hline 
\multicolumn{12}{c}{$d=10$}\tabularnewline
\hline 
MCRMHMC  & 1000 & 6.1 & 6.1 & 622 &  912 &  928 &  987  & 101 &  149  & 152 &  161\tabularnewline
Gibbs  &  5000000 & 2.3 & 2.4 & 588& 698  & 1295 & 1613 & 250 & 297 & 548 & 686\tabularnewline
EHMC  & 5000 & 6.3 & 6.5 & (8.9) & (48) & (20) & (161) & (1.4) & (7.3) & (3.2) & (25) \tabularnewline
Stan & 5000 & 4.0 & 5.4 & (4627) & (4839) & (4534) & (4856) & (692) & (918) & (704) & (920)\\
EigenRMHMC & 1000 & 1640 & 1694 & 17 & 56 & 32 & 76 & 0.01 & 0.03 & 0.02 & 0.04\\
\hline 
\multicolumn{12}{c}{$d=100$}\tabularnewline
\hline 
MCRMHMC  & 1000 & 154 &   156 &  482 & 628 & 398& 533 & 3.1 & 4.0 & 2.6 & 3.4\tabularnewline
Gibbs  &  5000000 & 20 & 20 & (42) & (76) & (111) & (243) & (2.1) & (3.8) & (5.5) & (12) \tabularnewline
EHMC  & 5000 &  89 & 91 & (5.4) & (11) & (8.4) & (33) & (0.06) & (0.12) & (0.09) & (0.36)\tabularnewline
Stan & 5000 & 33 & 42 & (4455) & (4742) & (4537) & (4844) & (92) & (114) & (96) & (116) \\
\hline 
\multicolumn{12}{c}{$d=1000$}\tabularnewline
\hline 
MCRMHMC  &     1000 &  11778 & 11927 & 331 & 636 & 596 & 908 & 0.03 & 0.05 & 0.05 & 0.08 \tabularnewline
Gibbs  & 5000000 & 198 & 198 &  (5.9) & (15) & (7) & (49) & (0.03) & (0.08) & (0.03) & (0.25)\tabularnewline
EHMC  & 5000 & 1686 & 1708 & (3.1) & (3.4) & (4.6) & (10) & (0.002) & (0.002) & (0.003) & (0.006)   \tabularnewline
Stan & 5000 & 487 & 511 & (3718) & (4391) & (3976) & (4776) & (6.8) & (8.7) & (6.7) & (9.5)\\
\hline 
\end{tabular}\caption{\label{tab:Results-for-the-funnel-ar1}Results for the funnel AR(1)
model (Equations \ref{eq:funnel_def_1}-\ref{eq:funnel_def_3}) experiment.
Each combination of method and $d\in\{10,100,1000\}$
was repeated 10 independent times and the numbers presented are minimum (min) and means (mean)
across these replica. Visual inspection
of output (Figures \ref{fig:Trace-plots-and-funnel-ar1_1} -
\ref{fig:Trace-plots-and-funnel-ar1_3}) and Kolmogorov-Smirnov tests 
(applied for MCRMHMC and Stan)
indicate that some samplers failed to explore the relevant support, and the corresponding
numbers are given in parentheses. The results for the Gibbs
sampler are only recorded every 1000th iteration, thus the maximum
ESSs for Gibbs are 5000.
Tuning parameters for MCRMHMC are: $d=10$: 
$u_{d}=\exp(2.0),$ $\varepsilon=0.3$, $l\sim U(30,40)$, $d=100$: 
$u_{d}=\exp(2.5),$ $\varepsilon=0.15$, $l\sim U(110,130)$ and $d=1000$:
$u_{d}=\exp(3.0),$ $\varepsilon=0.075$, $l\sim U(1100,1300)$. Tuning parameters
for EHMC are $d=10$: $\varepsilon=0.05,$ $l\sim U(1000,1500)$, 
$=100$: $\varepsilon=0.025$, $l\sim U(2000,3000)$ and $d=1000$: 
$\varepsilon=0.001$, $l\sim U(4000,6000)$. For Stan, default tuning parameters, 
except \texttt{adapt\_delta} = 0.999
were used. 
Tuning parameters for EigenRMHMC are $d=10$: $u=\exp(-1.0)$, $\varepsilon=0.3$ and
$l\sim U(200,300)$.}
\end{table*}
\begin{figure*}
\centering{}\includegraphics[scale=0.65]{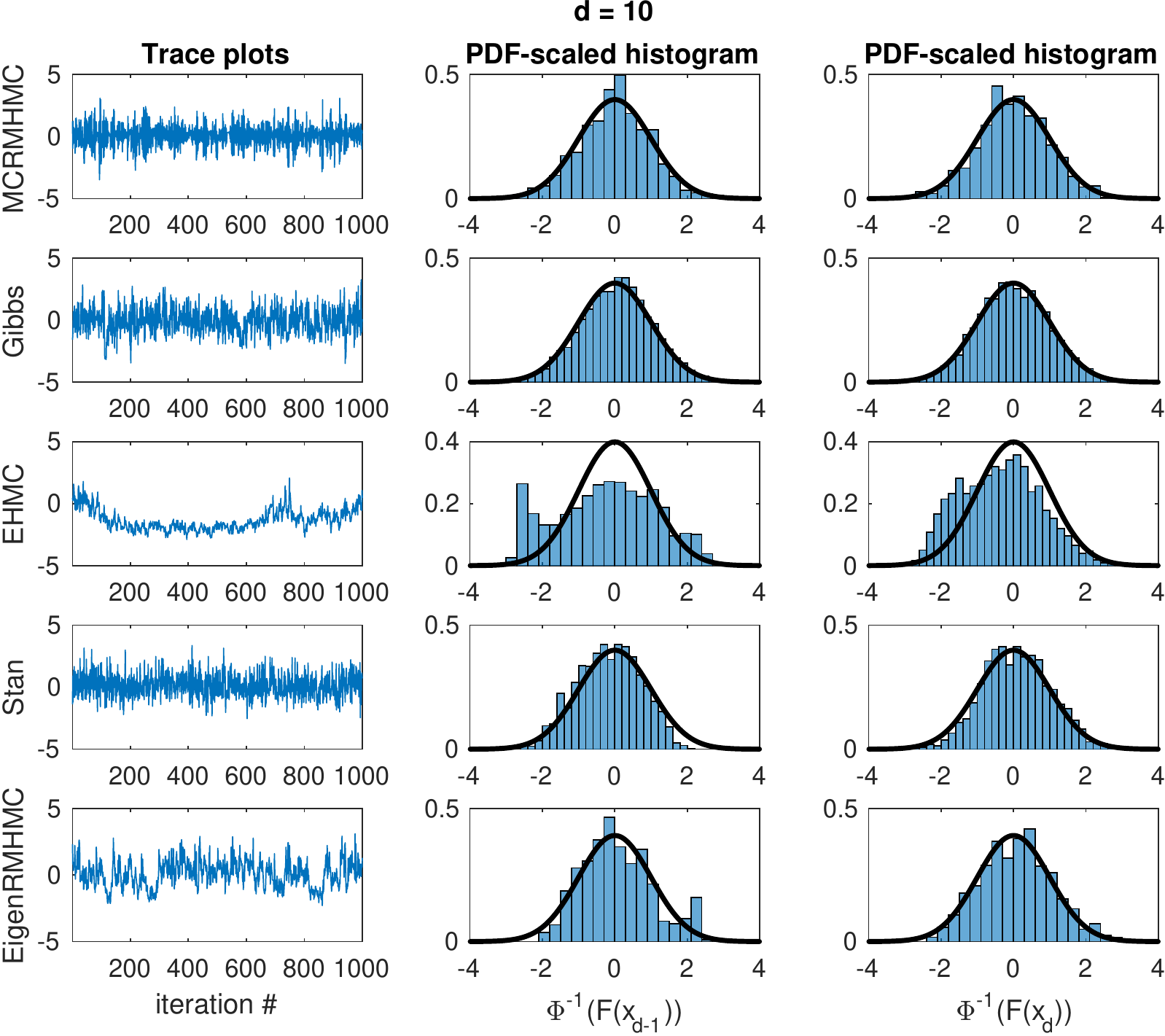}\caption{\label{fig:Trace-plots-and-funnel-ar1_1}
Trace plots and histograms
for the funnel AR(1) model experiment, $d=10$. Left column displays trace
plots of $\Phi^{-1}(F_{x_{d}}(x_{d}))$, which should have a standard
Gaussian distribution. The middle column displays density-scaled histograms
of $\Phi^{-1}(F_{x_{d-1}}(x_{d-1}))$, along with a standard Gaussian
density. The right column displays density-scaled histograms of $\Phi^{-1}(F_{x_{d}}(x_{d}))$
along with a standard Gaussian density. }
\end{figure*}
\begin{figure*}
\centering{}\includegraphics[scale=0.65]{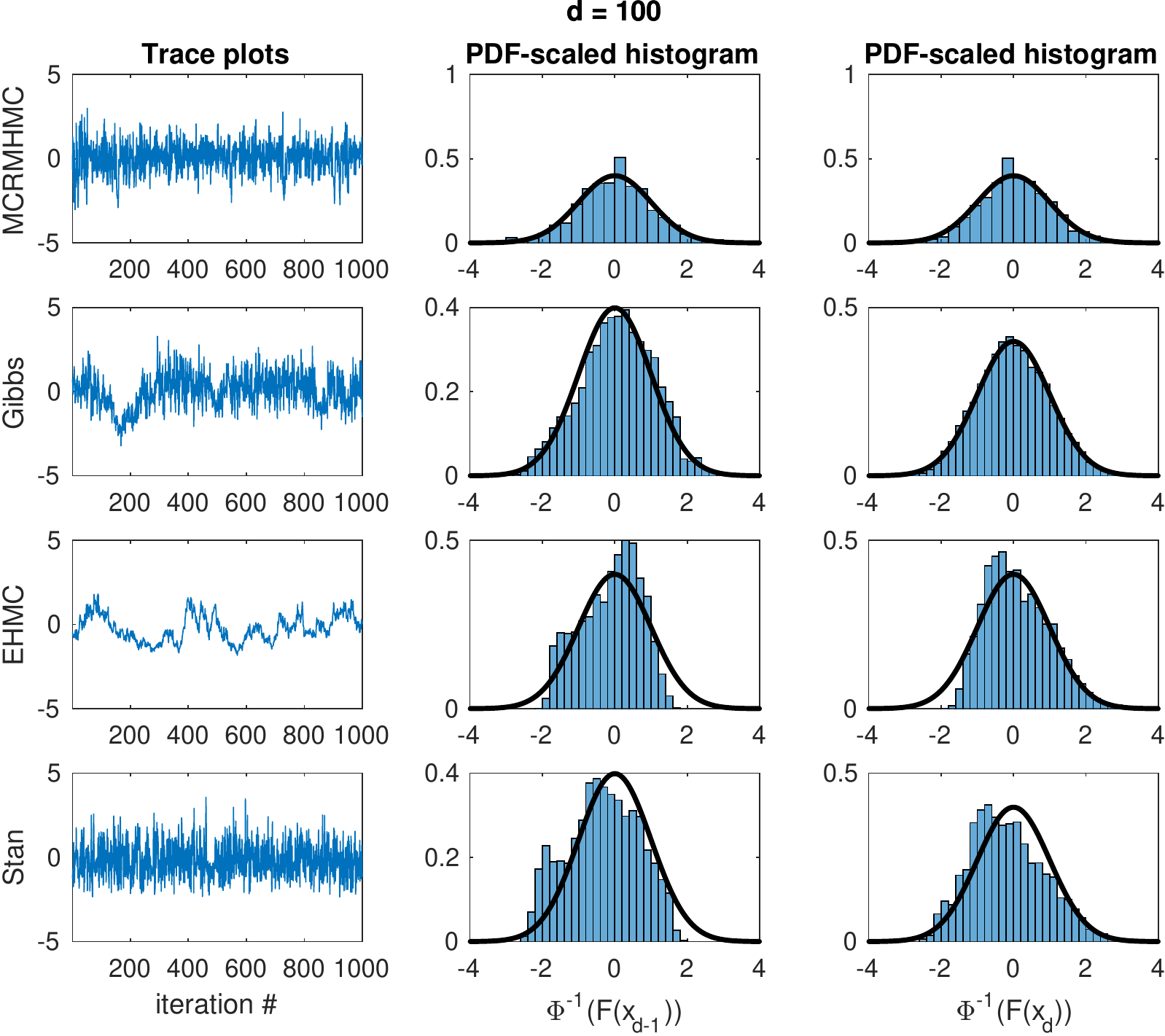}\caption{\label{fig:Trace-plots-and-funnel-ar1_2}
Trace plots and histograms
for the funnel AR(1) model experiment, $d=100$. Left column displays trace
plots of $\Phi^{-1}(F_{x_{d}}(x_{d}))$, which should have a standard
Gaussian distribution. The middle column displays density-scaled histograms
of $\Phi^{-1}(F_{x_{d-1}}(x_{d-1}))$, along with a standard Gaussian
density. The right column displays density-scaled histograms of $\Phi^{-1}(F_{x_{d}}(x_{d}))$
along with a standard Gaussian density. }
\end{figure*}
\begin{figure*}
\centering{}\includegraphics[scale=0.65]{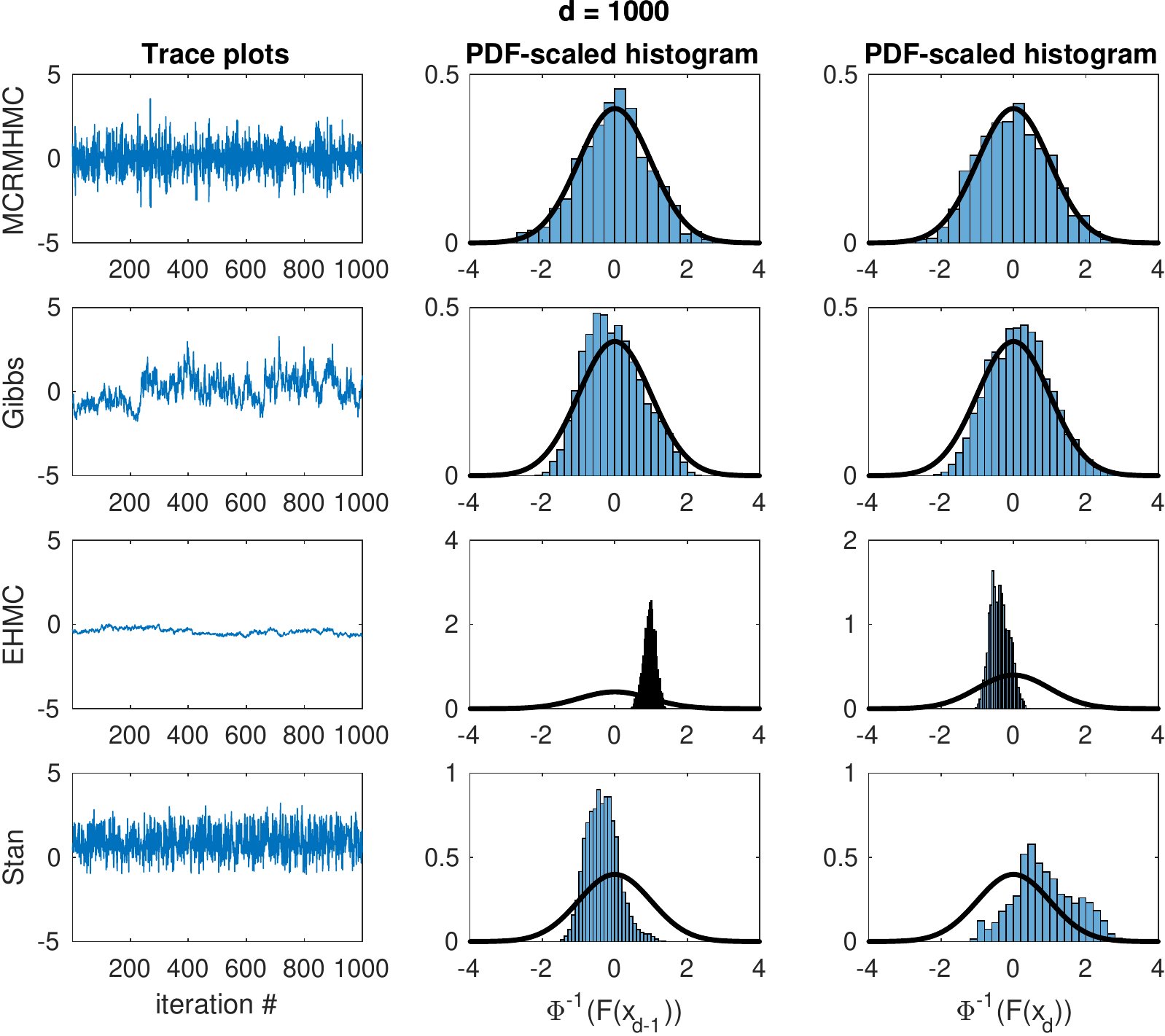}\caption{\label{fig:Trace-plots-and-funnel-ar1_3}
Trace plots and histograms
for the funnel AR(1) model experiment, $d=1000$. Left column displays trace
plots of $\Phi^{-1}(F_{x_{d}}(x_{d}))$, which should have a standard
Gaussian distribution. The middle column displays density-scaled histograms
of $\Phi^{-1}(F_{x_{d-1}}(x_{d-1}))$, along with a standard Gaussian
density. The right column displays density-scaled histograms of $\Phi^{-1}(F_{x_{d}}(x_{d}))$
along with a standard Gaussian density. }
\end{figure*}

Also here, four alternatives to the proposed methodology were considered,
namely Gibbs sampling, EHMC, Stan and EigenRMHMC (only for $d=10$). 
The model allows for a full set of
univariate conditional posteriors with standard distributions, which
are applied in the Gibbs sampler. For the Gibbs sampler, only every
1000th iteration was recorded due to the very slow mixing. Also here, the Euclidian
metric HMC was implemented using an identity mass matrix.
Due to the substantial ``funnel'' nature of the target, the integrator
step size for EHMC  must be chosen quite small to be able to explore
the low-variance region of the target, which in turn leads to very
high acceptance probabilities \citep{1212.4693}. However, finding
a reasonable distribution for $l$ that simultaneously leads 
to MCMC samples with low autocorrelation seems impossible. Also for Stan, small step sizes were imposed to account
for differences in scales 
via setting the acceptance rate target \texttt{adapt\_delta} equal to 0.999.
For MCRMHMC, again $\mathbf{x}_{1:d-1}|x_d$ is Gaussian which
suggest $K=d-1$, and the sparsity structure and tuning strategy
is identical as for the model in Section \ref{subsec:Twisted-Gaussian-mean}.
For EigenRMHMC, again the regularisation
parameter $u$ was chosen in order to produce few divergences in the
integrator fixed point iterations. However, for this $u$, it appears that
the regularisation interferes with representation of the AR(1) process precision, and
thus precludes finding small values of $l$ that produce high fidelity samples 
(see Section \ref{sec:chol_discussion}). The applied values of $l$ are larger than those
required for MCRMHMC.
The experimental setup is otherwise identical to that of Section
\ref{subsec:Twisted-Gaussian-mean}.

Due to the simple structure of the target, it is clear that the marginal
cumulative distribution functions of $x_{i}$, say $F_{x_{i}}$, are
easily determined from $\exp(x_{d})\sim\text{gamma}(1,0.1)$ and $\sqrt{0.1(1-0.999^{2})}x_{i}\sim t_{2}$,\\
 $i=1,\dots,d-1$.
This information can be exploited to determine the quality of the
convergence for the different methods. Results from the simulation
experiment for $d\in\{10,100,1000\}$ are presented in Table \ref{tab:Results-for-the-funnel-ar1}
and Figures \ref{fig:Trace-plots-and-funnel-ar1_1} - \ref{fig:Trace-plots-and-funnel-ar1_3}. 
Looking first at
Figures \ref{fig:Trace-plots-and-funnel-ar1_1} - \ref{fig:Trace-plots-and-funnel-ar1_3}, 
where the middle and
right columns present density-scaled histograms of the MCMC output
for a representative replica, transformed to have standard Gaussian
marginal distribution via the $\Phi^{-1}\circ F_{x_{i}}$ transformation
for $x_{d-1}$ and $x_{d}$ respectively. Firstly, it is seen that
EHMC, even with very large $E(l)$ and very high acceptance rates,
is unable to fully explore the target for all considered $d$s for
this model. Secondly, this appears also to be the case for the Gibbs
sampler for $d=100,1000$. 
Also Stan fails to explore the target fully (most easily seen in 
Figures \ref{fig:Trace-plots-and-funnel-ar1_2}, \ref{fig:Trace-plots-and-funnel-ar1_3}), 
and Kolmogorov-Smirnov
tests reject that the Stan MCMC samples come from the true marginals in all cases.
From Table \ref{tab:Results-for-the-funnel-ar1}
it is seen that MCRMHMC produces high-fidelity samples in all cases,
but that in the $d=10$ case, Gibbs sampling produces effective samples
faster than MCRMHMC. EigenRMHMC is four orders of magnitude slower than MCRMHMC and Gibbs in
the $d=10$ case.

Overall, the simulation studies for models (\ref{eq:twisted-ar-1}-\ref{eq:twisted-ar-3})
and (\ref{eq:funnel_def_1}-\ref{eq:funnel_def_3}) indicate that
MCRMHMC is capable of producing reliable results for challenging models
with diverse forms of non-linearities and strong dependencies structures.
On the other hand, the contending methods may require prohibitively long
MCMC chains to accurately represent the target distribution.
Having said that, one can argue that the conditional Gaussian AR(1)
models in (\ref{eq:twisted-ar-1}-\ref{eq:twisted-ar-3}) and (\ref{eq:funnel_def_1}-\ref{eq:funnel_def_3})
pose too small challenges for MCRMHMC. Therefore, also
a non-linear state space model where the prior for the latent variables
is far from jointly Gaussian is considered. 

\section{Application to a state space model\label{sec:Application-to-a}}

In this section, a Euler-Maruyama-discretized version
of the \citet{chan_et_al_1992} constant elasticity of variance model for
short term interest rates, observed with Gaussian noise, is considered.
The model is fully specified by 
\begin{equation}
y_{t} =x_{t}+\sigma_{y}\eta_{t},\;\eta_{t}\sim i.i.d.\;N(0,1),\;t=1,\dots,T,\label{eq:CEV_1}
\end{equation}
\begin{multline}
x_{t}  =x_{t-1}+\Delta(\alpha-\beta x_{t-1})+\sigma_{x}\sqrt{\Delta}x_{t-1}^{\gamma}\varepsilon_{t},\\
\;\varepsilon_{t}\sim i.i.d.\;N(0,1),\;t=2,\dots,T\label{eq:CEV_2}
\end{multline}
\begin{equation}
x_{1} \sim N(0.09569,0.01^{2}),\label{eq:CEV_3}
\end{equation}
where $\Delta=1/252$ correspond to a daily sampling frequency for
a yearly time scale. Here $\mathbf{x}_{1:T}$ models an unobserved
short term interest rate, which is observed with noise in $\mathbf{y}_{1:T}$.
The prior mean of $x_{1}$ is set equal to the first observation $y_{1}$.
Due to the non-constant volatility term in the $x_{t}$ process, the
joint prior for $\mathbf{x}_{1:T}$ deviates strongly from being Gaussian.
The model is completed with the following, (improper) prior assumptions
and transformations to prepare for MCRMHMC sampling:
\begin{align*}
x_{T+1} & =\alpha\sim N\left(0,\frac{100}{\Delta^{2}}\right),\\
x_{T+2} & =\beta\sim N\left(\frac{1}{\Delta},\frac{100}{\Delta^{2}}\right),\\
x_{T+3} & =\log(\sigma_{x}^{2})\sim\text{uniform}(-\infty,\infty),\\
x_{T+4} & =\gamma\sim\text{uniform}(0,\infty),\\
x_{T+5} & =\log(\sigma_{y}^{2})\sim\text{uniform}(-\infty,\infty),
\end{align*}
(the priors for $\alpha,\beta$ correspond to $N(0,100)$ priors on
$\alpha^{\prime},\beta^{\prime}$ in the parameterization $E(x_{t}|x_{t-1})=\alpha^{\prime}+\beta^{\prime}x_{t-1}$).
The dataset considered was $T=3082$ observations of 7-day Eurodollar
deposit spot rates from January 2, 1983 to February 25, 1995. The
dataset has previously been used in \citet{aitsahalia1996} and, in
the context of model (\ref{eq:CEV_1}-\ref{eq:CEV_3}), in \citet{1601.01125}.
To implement MCRMHMC with $d=T+5$, it is first observed that even though
$p(x_{1:T}|x_{T+1:T+5})$ is not uniformly log-concave, it appears
that $K=T+2$ is a suitable choice since the observations are highly informative
with respect to the latent process, and further that $\alpha,\beta$
interacts only linearly with $\mathbf{x}_{1:T}$. The remaining active
regularisation parameters were taken to be $u_{T+3}=\exp(6.0)$,
$u_{T+4}=\exp(0.0)$, $u_{T+5}=\exp(6.0)$, which were determined
using the heuristic detailed in Section \ref{subsec:Implementation-and-tuning}.
Moreover, $l\sim U(90,180)$ and $\varepsilon=0.03$ with 15\%
jittering of step size were used, which correspond to integration times ranging
between $\approx2.3$ and $\approx6.2$. The applications of MCRMHMC
were done using 1100 MCMC iterations where the 100 first iterations
were discarded as warm up.
\begin{table*}
\centering{}%
\begin{tabular}{clccccc}
\hline 
 &  & \multicolumn{2}{c}{MCRMHMC} &  & \multicolumn{2}{c}{Particle Gibbs}\tabularnewline
\hline 
 & CPU time (hours) & \multicolumn{2}{c}{4.5} &  & \multicolumn{2}{c}{2.0}\tabularnewline
 & \# MCMC iterations & \multicolumn{2}{c}{1000+100} &  & \multicolumn{2}{c}{80000+20000}\tabularnewline
 & across replica & min & mean & & min & mean \\
\hline 
 &  &  &  & \tabularnewline
$\alpha$ & Post. mean & & 0.0099 &  & & 0.0098\tabularnewline
 & Post. std. & & 0.0088 &  & & 0.0089\tabularnewline
 & ESS &       1000 &   1000 &  &  36472 & 60801\tabularnewline
 & ESS/hour CPU time  & 222 & 224 &  &  17911 & 29755\tabularnewline
 &  &  &  & \tabularnewline
$\beta$ & Post. mean & & 0.168 &  &    &   0.168\tabularnewline
 & Post. std. & & 0.172 &  &  &0.173\tabularnewline
 & ESS &        1000 & 1000 &  &    56201 &     70998\tabularnewline
 & ESS/hour CPU time & 222 & 224 &  &  27600 &       34752\tabularnewline
 &  &  &  & \tabularnewline
$\sigma_{x}$ & Post. mean  & &        0.41 &  & & 0.41\tabularnewline
 & Post. std. & & 0.06 &  & & 0.06\tabularnewline
 & ESS & 405 & 579 &  & 52 &  79\tabularnewline
 & ESS/hour CPU time & 90 & 130 &  & 26 & 38\tabularnewline
 &  &  &  & \tabularnewline
$\gamma$ & Post. mean & & 1.18 &  &   &      1.19\tabularnewline
 & Post. std. & &     0.06 &  &    &  0.06\tabularnewline
 & ESS &   423 &  564 &  &  53 & 78\tabularnewline
 & ESS/hour CPU time &  94 & 127 &  & 26 & 38\tabularnewline
 &  &  &  & \tabularnewline
$\sigma_{y}$ & Post. mean &  &   0.00054 &  & &  0.00054\tabularnewline
 & Post. std. & & 2.3e-5 & & & 2.2e-5\tabularnewline
 & ESS & 495 & 582 &  &    409   &  557\tabularnewline
 & ESS/hour CPU time & 110 & 131 &  & 201 & 272\tabularnewline
 &  &  &  & \tabularnewline
$x_{1}$ & Post. mean & & 0.095 &  & & 0.095\tabularnewline
 & Post. std. & & 0.0005 &  & & 0.0005\tabularnewline
 & ESS &  1000 &  1000 &  &   71512 & 73304\tabularnewline
 & ESS/hour CPU time & 222 & 224 &  & 35087 & 35881\tabularnewline
 &  &  &  & \tabularnewline
$x_{T}$ & Post. mean & &      0.061 &  & &      0.061\tabularnewline
 & Post. std. & &   0.0005 &  &  &  0.0005\tabularnewline
 & ESS &  1000 & 1000 &  &  72415 & 74206\tabularnewline
 & ESS/hour CPU time & 222 & 224 &  & 35519 & 36324\tabularnewline
 &  &  &  & \tabularnewline
\hline 
\end{tabular}\caption{\label{tab:Results-for-Bayesian-CEV}Results for Bayesian inference
for model (\ref{eq:CEV_1}-\ref{eq:CEV_3}) applied to a data set
of Eurodollar interest rates with $T=3082$. Posterior mean and posterior
standard deviation are denoted by Post. mean and Post. std. respectively.
The numbers reported are the minimum (min) and  mean (mean) across 10 independent replications
of the experiments. }
\end{table*}
\begin{figure*}
\begin{centering}
\includegraphics[scale=0.7]{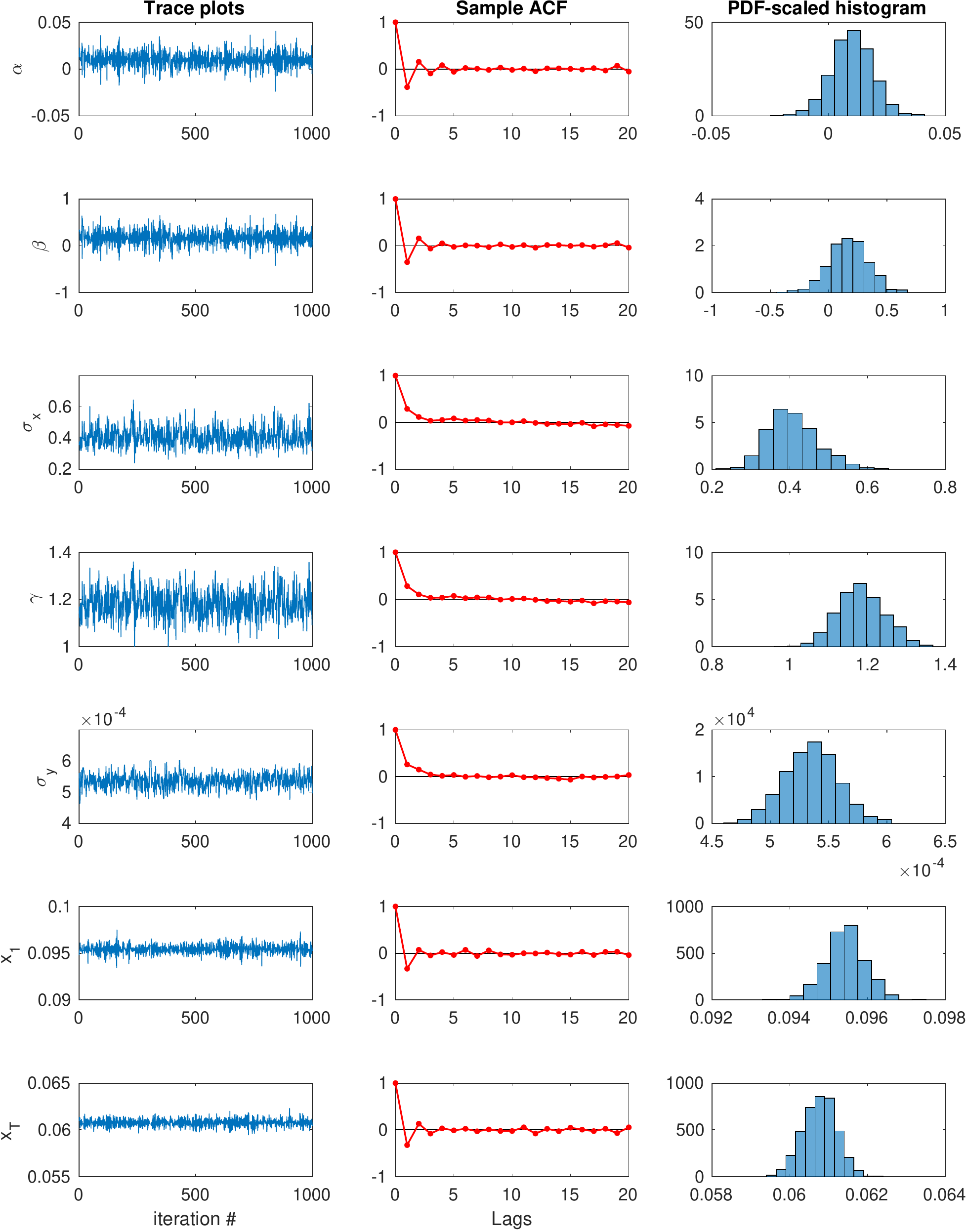}\caption{\label{fig:Trace-plots,-Sample-MCRMHMC}Trace plots, Sample autocorrelation
functions and histograms for the CEV model (\ref{eq:CEV_1}-\ref{eq:CEV_3})
based on MCRMHMC. The reported plots are for a representative replica,
and depicts only the 1000 post warm up samples.}
\par\end{centering}
\end{figure*}
\begin{figure*}
\centering{}\includegraphics[scale=0.7]{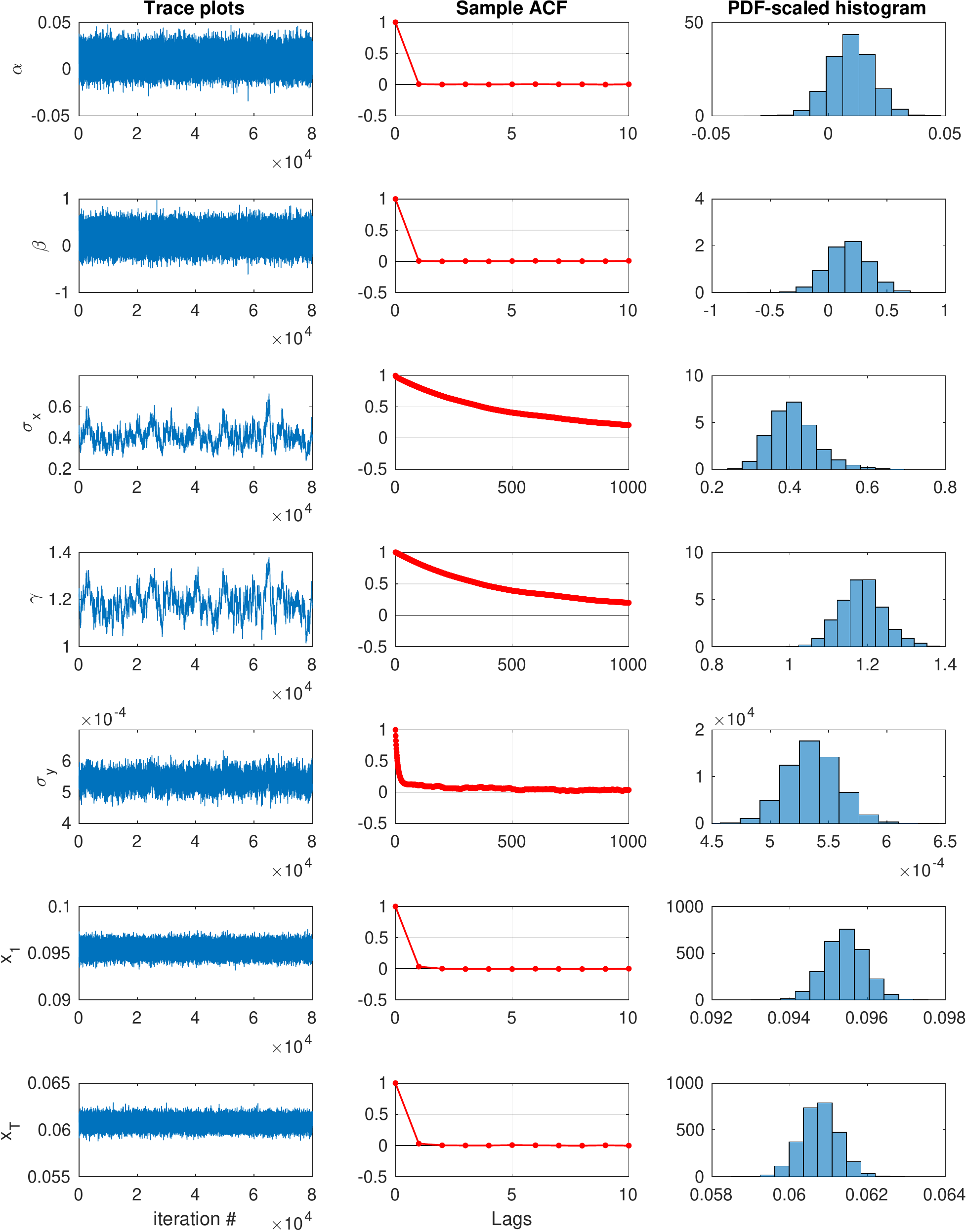}\caption{\label{fig:Trace-plots,-Sample-PG}Trace plots, Sample autocorrelation
functions and histograms for the CEV model (\ref{eq:CEV_1}-\ref{eq:CEV_3})
based on Particle Gibbs. The reported plots are for a representative
replica, and depicts only the 80000 post warm up samples.}
\end{figure*}

As a reference for the proposed methodology, a Particle
Gibbs sampler \citep{RSSB:RSSB736} using the Particle EIS filter
\citep{Scharth2016133} with Ancestor Sampling \citep{JMLR:v15:lindsten14a} was used.
This methodology is discussed in detail in \citet{1601.01125} and
should be regarded as a ``state of the art'' Gibbs sampling procedure
for state space models where the latent state ($x_{1:T}$ in current
notation) is updated in a single Gibbs block with close to perfect
mixing. The Gibbs sampler was implemented fully
in C++, and with a Gaussian random walk MH updating mechanism for
$\gamma$ (with proposal standard deviation 0.025 corresponding to
acceptance rates of 20-30\%), but is otherwise identical to the one
described in \citet{1601.01125}. A total of 100000 Gibbs iterations
were performed, with the first 20000 iterations discarded as warm
up. All experiments in this section were run on a 2016 iMac with a
3,1 GHz Intel Core i5 processor and 8 Gb of RAM, and each experiment
was repeated 10 times using different random number seeds. 

It is well known that Gibbs sampling for latent variable models often
lead to poor mixing for the parameters determining the volatility
of the latent process. This is at least partly due to ``funnel''-like
non-linear dependencies between the relevant parameters and the complete
latent state. Thus, of particular interest is whether MCRMHMC is able
to improve the sampling of these parameters by including both parameters
and latents in the same updating block. 

Results for MCRMHMC and Particle
Gibbs sampling for the CEV model are presented in Table \ref{tab:Results-for-Bayesian-CEV}
and Figures \ref{fig:Trace-plots,-Sample-MCRMHMC} and \ref{fig:Trace-plots,-Sample-PG}.
It is seen from Table \ref{tab:Results-for-Bayesian-CEV} that both
methods produce close to perfect sampling for parameters $\alpha,\;\beta,$
$x_{1},\;x_{T}$, and indeed also the other latents (not reported).
However, for $\sigma_{x},\;\gamma$ and to some degree $\sigma_{y}$,
the performance of the Particle Gibbs sampler deteriorates substantially,
and for $\sigma_{x}$ and $\gamma$, MCRMHMC produces effective samples
at a faster rate than the Particle Gibbs (both in mean and minimum across the replica). 
Thus, looking at the minimum
(across $d$ dimensions) ESS per computing time, MCRMHMC performs
better than Particle Gibbs, even if the Particle Gibbs updating mechanism for $\mathbf x_{1:T}$
is both fast and produces close to independent updates. 

From Figure \ref{fig:Trace-plots,-Sample-MCRMHMC}, it is seen that
the samples from MCRMHMC explores the relevant support extremely fast
for all dimensions, and it appears that only a few hundred iterations
would be sufficient to produce a good representation of the joint posterior.
A further, interesting artefact, is that for the conditionally log-concave
latents/parameters ($\mathbf{x}_{1:T+2}$), the applied distributions
for $l$ and $\varepsilon$ produce mildly negatively autocorrelated
MCMC iterations, whereas there is a small positively correlated dependence
for the parameters in need of regularisation ($\mathbf{x}_{T+3:T+5}$).
This indicates that there is potentially scope for further improvement
of the tuning of sampler, where it should be possible to align better
the integration times needed to produce close to zero autocorrelation
in all dimensions at the same time by a more refined selection of
$\mathbf{u}_{T+3:T+5}$. 

From Figure \ref{fig:Trace-plots,-Sample-PG}, the poor performance
of Particle Gibbs for the volatility-determining parameters $\sigma_{x}$,
$\gamma$ is clearly seen from the trace plots and autocorrelation
functions, and it is not mitigated by extremely fast mixing of the latents
and the mean-structure of (\ref{eq:CEV_2}). Rather, the poor mixing is an artefact
of handling the model in Gibbs manner. Moreover, it is seen from the
trace plots that $\sigma_{x}$ and $\gamma$ are strongly correlated
a-posteriori ($\text{corr}(x_{T+3},x_{T+4})\simeq0.99$) which is
handled very well by MCRMHMC, but poses problems for Gibbs sampling.

\section{Discussion}

This paper has presented a modified Cholesky factorization suitable
for turning the log-target Hessian matrix into a suitable metric tensor
for Riemann manifold Hamiltonian Monte Carlo. The resulting modified
Cholesky RMHMC is shown, both in a simulation study and in a real data
experiment, to be competitive, in particular for high-dimensional,
challenging sampling problems. The method is in particular well
suited for sampling problems where the Hessian is sparse, as unlike
methods based spectral decomposition, the modified Cholesky factorization
can exploit sparsity, which holds the potential for speeding up computations
drastically. Moreover, the method can exploit that the conditional
posterior of some subset of the state vector $\mathbf{x}$ is log-concave,
e.g. latents conditional on scale \citep{shepard_pitt97,RSSB:RSSB700},
and thereby forego unnecessary regularisation that slows down exploration
of the target.

This paper focusses in particular on how to turn potentially indefinite
Hessian matrices into useful scaling information. However, to make
the proposed methodology more automatic and thus be suitable for inclusion
general purpose softwares (e.g. Stan), further work is in order.
In particular, the methodology would likely benefit from dynamic selection
of integration times \citep{JMLR:v15:hoffman14a,1601.00225} and automatic
selection of integration step sizes \citep[e.g.][]{JMLR:v15:hoffman14a}.
Also a more refined and automated tuning of the active regularisation parameters during warm up,
which takes into account dependencies in the target, holds scope for
further work. One potential such direction is to measure the sensitivity
with respect to the regularisation parameters on the first iterate
of the fixed point iterations for (\ref{eq:implicit_1}-\ref{eq:implicit_2}).
Finally, internal reordering of the variables and selection of $K$
may also be done more automatically provided that a robust selection of the active 
regularisation parameters is in place, as one may put the variables requiring
the largest regularisation parameters last in $\mathbf x$, whereas
those not requiring regularisation could be put first in $\mathbf x$ and
$K$ could be chosen accordingly. However, such a reordering
must be weighed against the impact it has on exploitable sparsity 
of $L(\mathbf x)$, and therefore optimal, automatic selection of $K$ and reordering
of the variables for a general target also holds scope for further work.

In addition, to make the methodology more user friendly, 
some form of sparse automatic differentiation for calculating
the Hessian of the log-target should be implemented, while retaining automatic differentiation
of the Hamiltonian with respect to $\mathbf{x}$. Interestingly, these
calculations are similar to the practice of differentiating Laplace
approximations (for integrating out latents) with respect to parameters
\citep{skaug_fournier1996aam,JSSv070i05}. In particular, at least
two avenues must be explored: 1) Apply first order backward AD
to both the modified Cholesky code and the second order AD code for the Hessian of the
log-target (by differentiating the internal Hessian AD computations). 2) Directly compute gradient, sparse Hessian and 3rd order
sparse derivative tensor of the log-target using AD, and combine these
and the differential of the modified Cholesky factorization to find
$\nabla_{\mathbf{x}}H$ similarly to \citet{1212.4693}. Finding the
best method among these requires further work. 

\bibliographystyle{chicago}
\bibliography{kleppe}

\end{document}